\documentclass[traditabstract]{aa}  

\usepackage{lineno}
\usepackage{graphicx}
\usepackage{txfonts}
\usepackage{hyperref}
\usepackage{natbib}
\begin{document} 

   \title{Hunting for the nature of the enigmatic narrow-line Seyfert 1 galaxy PKS 2004-447}

   \author{M. Berton\inst{1,2}\thanks{e-mail: marco.berton@utu.fi}
   	\and G. Peluso\inst{3,4}
   	\and P. Marziani\inst{3}
	\and S. Komossa\inst{5}
	\and L. Foschini\inst{6}
   	\and S. Ciroi\inst{4}
   	\and S. Chen\inst{7}
	\and E. Congiu\inst{8}
	\and L.~C. Gallo\inst{9}
   	\and I. Bj\"orklund\inst{2,10}
	\and L. Crepaldi\inst{4}
	\and F. Di Mille\inst{11}
    \and E. J\"arvel\"a\inst{12}
	\and J. Kotilainen\inst{1}
	\and A. Kreikenbohm\inst{13,14}
	\and N. Morrell\inst{11}
	\and P. Romano\inst{6}
	\and E. Sani\inst{15}
	\and G. Terreran\inst{16}
	\and M. Tornikoski\inst{2,10} 
	\and S. Vercellone\inst{6}
	\and A. Vietri\inst{4}
          }

   \institute{
   	$^1$ Finnish Centre for Astronomy with ESO (FINCA), University of Turku, Vesilinnantie 5, FI-20014 University of Turku, Finland; \\
	$^2$ Aalto University Mets{\"a}hovi Radio Observatory, Mets{\"a}hovintie 114, FI-02540 Kylm{\"a}l{\"a}, Finland; \\
	$^3$ INAF - Osservatorio Astronomico di Padova, Vicolo dell'Osservatorio 5, 35122 Padova, Italy; \\
	$^4$ Dipartimento di Fisica e Astronomia "G. Galilei", Universit\`a di Padova, Vicolo dell'Osservatorio 3, 35122 Padova, Italy; \\
	$^{5}$ Max-Planck-Institut f\"ur Radioastronomie, Auf dem H\"ugel 69, 53121 Bonn, Germany; \\
	$^6$ INAF - Osservatorio Astronomico di Brera, via E. Bianchi 46, 23807 Merate (LC), Italy;\\
	$^{7}$ Department of Physics, Technion 32000, Haifa 32000, Israel; \\
	$^{8}$ Departamento de Astronom\'{i}a, Universidad de Chile, Camino del Observatorio 1515, Las Condes, Santiago, Chile; \\	
	$^9$ Department of Astronomy and Physics, Saint Mary's University, 923 Robie Street, Halifax, NS B3H 3C3, Canada; \\
	$^{10}$ Aalto University Department of Electronics and Nanoengineering, P.O. Box 15500, FI-00076, Aalto, Finland; \\
	$^{11}$ Las Campanas Observatory, Carnegie Observatories, Colina El Pino, Casilla 601, La Serena, Chile; \\
	$^{12}$ European Space Agency (ESA), European Space Astronomy Centre (ESAC), Camino Bajo del Castillo s/n, 28692 Villanueva de la Ca\~nada, Madrid, Spain; \\
	$^{13}$ Dr. Karl Remeis Observatory, Universit\"at Erlangen-N\"urnberg, Sternwartstr. 7, D-96049 Bamberg, Germany; \\
	$^{14}$ Lehrstuhl fur Astronomie, Universit\"at W\"urzburg, Campus Hubland Nord, Emil-Fischer-Strasse 31, D-97074 W\"urzburg, Germany; \\
	$^{15}$ European Southern Observatory, Alonso de Cordova 3107, Casilla 19, Santiago 19001, Chile; \\
	$^{16}$ Center for Interdisciplinary Exploration and Research in Astrophysics (CIERA), Department of Physics and Astronomy, Northwestern University, Evanston, IL 60208, USA. \\
             }

   \date{Received \today; }
   
   \authorrunning{M. Berton et al.}
\titlerunning{Hunting the nature of PKS 2004-447}

  \abstract{Narrow-line Seyfert 1 (NLS1) galaxies are a class of active galactic nuclei (AGN) that, in some cases, can harbor powerful relativistic jets. One of them, PKS 2004-447, shows $\gamma$-ray emission, and underwent its first recorded multifrequency flare in 2019. However, past studies revealed that in radio this source can be classified as a compact steep-spectrum source (CSS), suggesting that, unlike other $\gamma$-ray sources, the relativistic jets of PKS 2004-447 have a large inclination with respect to the line of sight. We present here a set of spectroscopic observations of this object, aimed at carefully measuring its black hole mass and Eddington ratio, determining the properties of its emission lines, and characterizing its long term variability. We find that the black hole mass is $(1.5\pm0.2)\times10^7$ M$_\odot$, and the Eddington ratio is 0.08. Both values are within the typical range of NLS1s. The spectra also suggest that the 2019 flare was caused mainly by the relativistic jet, while the accretion disk played a minor role during the event. In conclusion, we confirm that PKS 2004-447 is one of the rare examples of $\gamma$-ray emitting CSS/NLS1s hybrid, and that these two classes of objects are likely connected in the framework of AGN evolution. }


   \maketitle

\newcommand{\kms}{km s$^{-1}$}
\newcommand{\ergs}{erg s$^{-1}$}
\newcommand{\chired}{$\chi^2_\nu$}
\section{Introduction}
Since the launch of the \textit{Fermi} Satellite, narrow-line Seyfert 1 galaxies (NLS1s) have been identified as the third class of active galactic nuclei (AGN) that can harbor powerful beamed relativistic jets and produce $\gamma$-ray emission beside the two well-known classes of blazars, BL Lacertae objects (BL Lacs) and flat-spectrum radio quasars (FSRQs, \citealp{Abdo09a, Abdo09b, Abdo09c, Foschini10}). NLS1s are characterized by a relatively low full-width at half maximum (FWHM) of H$\beta$, which by definition must be lower than 2000 \kms, by a flux ratio [O III]/H$\beta <$ 3, and two bumps of Fe II multiplets, which indicate that these objects are type 1 AGN with an unobscured view of their central engine \citep{Osterbrock85, Goodrich89}. \par
The narrowness of the permitted lines observed in NLS1s is typically interpreted as a sign of low rotational velocity around a relatively low-mass black hole ($10^6-10^8$ M$_\sun$, \citealp{Boller96, Peterson00, Peterson11, Cracco16, Rakshit17a, Chen18}). The black hole is accreting close to or above the Eddington limit \citep{Boroson92, Sulentic00}, especially in the strongest Fe II emitters \citep{Du16}. Non-jetted NLS1s are typically hosted by a spiral galaxy with a pseudobulge \citep{Crenshaw03, Deo06, Orbandexivry11, Mathur12}. This ensemble of properties has led several authors to hypothesize that these NLS1s may represent an early evolutionary stage in the life of AGN, that will eventually grow into classical broad-line Seyfert 1 galaxies \citep{Grupe00, Mathur00, Sulentic00, Komossa06, Fraixburnet17a}. 
Jetted NLS1s seem to behave like their non-jetted counterparts. They are the low-mass tail of the FSRQ distribution \citep{Abdo09a, Abdo09c, Foschini15, Berton16a}, which in turn suggests that they may be the progenitors of FSRQs \citep{Berton16b}. NLS1s with misaligned relativistic jets may eventually evolve to form the parent population of FSRQs, i.e. high-excitation radio galaxies (HERG, \citealp{Berton16b, Berton17, Foschini17}). Several authors pointed out that a possible link between jetted NLS1s and other classes of young radio galaxies \citep[see][for a review]{Odea21}, such as compact steep-spectrum sources (CSS) and gigahertz-peaked sources (GPS), may exist \citep{Oshlack01, Gallo06a, Komossa06, Yuan08, Wu09, Caccianiga14, Caccianiga17, Schulz15, Gu15, Sulentic15b, Liao20, Zhang20, Odea21, Yao21}. It is also worth noting that in the X-rays there is even evidence that the corona in some non-jetted NLS1s exhibit jet-like properties like collimation and outflow. This behavior is remarkable, as it could indicate potential jet-like behavior even among non-jetted AGN \citep[see][for a review]{Gallo18}. \par
NLS1s are located on the horizontal branch of the so-called quasar main sequence (MS, \citealp{Sulentic15, Marziani18a}). The MS is the locus on the plane defined by the flux ratio between the Fe II multiplets and H$\beta$, known as R4570, and the FWHM(H$\beta$), where all type 1 AGN lie. This sequence was originally identified by means of principal component analysis \citep{Boroson92}. The MS can be roughly divided into two distinct populations of sources, called population A and B. Population A forms a horizontal branch in the MS, since its sources have different values of R4570 but they all have FWHM$\leq$4000 \kms. Population A can be binned into four groups, from A1 to A4, with values of R4570 increasing by 0.5 from group to group. Population B, instead, forms a vertical branch, since all of them have R4570$<$0.5 but FWHM$>$4000 \kms \citep{Sulentic00}. All NLS1s, because of their definition, belong to population A. \par
The main driver of the MS may be a decreasing Eddington ratio from population A to B \citep{Boroson92}. Some authors even hypothesized that the MS could be an evolutionary path for AGN \citep{Fraixburnet17a}. However, also some other factors seem to play a role in the MS, such as metallicity and inclination with respect to the line of sight \citep{Shen14, Panda19, Sniegowska21}. This last parameter can be particularly confusing, since in the presence of a flattened broad-line region (BLR) observed pole-on, a low inclination can produce narrow permitted lines even in the presence of large black hole masses due to the lack of Doppler broadening \citep{Decarli08}. Some authors suggested that this may be the case for jetted NLS1s \citep[e.g.,][]{Calderone13, Baldi16, Dammando19b}. However, recent observations dedicated to the host galaxies of these objects showed that jetted NLS1s are typically hosted in disk galaxies similarly to their non-jetted counterparts, and confirmed that their black hole mass is lower than that of FSRQs \citep[][but see \citealp{Dammando17, Dammando18}]{Anton08, Orbandexivry11, Mathur12, Leontavares14, Kotilainen16, Olguiniglesias17, Jarvela18, Berton19a, Olguiniglesias20, Hamilton21}. \par
The number of currently known $\gamma$-ray emitting NLS1s is rather limited, with approximately twenty sources classified to date \citep[e.g.,][see the review by \citealp{Komossa18}]{Romano18, Paliya19a, Jarvela20, Rakshit21b}. Because of their relatively high redshift (all but 3 have z$>$0.2), $\gamma$-NLS1s are rather faint in optical bands, and very few studies have been dedicated specifically to their optical spectra \citep[e.g.,][]{Komossa18a, Kynoch18, Kynoch19, Yao21}. Here we present new spectroscopic data for the southernmost (to date) $\gamma$-NLS1, PKS 2004-447. This NLS1 was identified as a $\gamma$-ray source soon after the launch of the \textit{Fermi} Satellite \citep{Abdo09c}, and at the end of 2019 it underwent its first $\gamma$-ray flare ever recorded \citep{Gokus19, Gokus21}. The goal of this work is to study its long-term behavior and the nature of this flare by means of optical spectroscopy. We also accurately measure some of its most important physical parameters such as the black hole mass, the Eddington ratio, and the emission line properties, and we determine its role in the family of $\gamma$-ray emitting NLS1s. In Sect.~2 we provide a brief review about the target of this paper, in Sect.~3 we describe the process of data reduction, in Sect.~4 we analyze the profiles of the most prominent emission lines, in Sect.~5 we determine its black hole mass, in Sect.~6 we study its time variability, and finally in Sect.~7 we provide a summary of our results. Although we are aware of the tension in the value of the Hubble constant \citep{Riess2019}, which would require the use of $H_0\sim 74$~\kms\ Mpc$^{-1}$ for the sources in the nearby Universe, throughout this work we adopt a standard $\rm \Lambda CDM$ cosmology, with a Hubble constant $H_0 = 70$ \kms\ Mpc$^{-1}$, and $\Omega_\Lambda = 0.73$ \citep{Komatsu11} to allow an easier comparison with previous works. 

\section{PKS 2004-447}
The target of this study is the $\gamma$-ray emitting NLS1 PKS 2004-447 (R.A. 20h 07m 55s, Dec. -44d 34m 44s, z = 0.240). Like most NLS1s, it is hosted in a spiral galaxy with a pseudobulge \citep{Kotilainen16}. The source was noticed early on due to its prominent radio emission, and it was originally included in the Parkes Half-Jansky Flat-Spectrum Sample because of its flat radio spectrum\footnote{Conventionally, a radio spectrum is defined as flat when $\alpha_\nu < 0.5$, and steep when $\alpha_\nu > 0.5$.} between 2.7 and 5.0 GHz (F$_{2.7\;\rm{GHz}}$ = 0.81 Jy, $\alpha_\nu$ = 0.36, with F$_\nu \propto \nu^{-\alpha}$, \citealp{Drinkwater97}). However, this radio spectral measurement was carried out on non-simultaneous data. Later observations revealed instead a steep spectral index of $\alpha_\nu = 0.67$, suggestive of a radio classification as a CSS \citep{Oshlack01, Gallo06a}. Such result was confirmed more recently with new high-resolution observations in radio, that additionally found a core-jet morphology, and a flattening in the spectrum below 2 GHz \citep{Schulz15}. This may be interpreted as a turnover, which is a defining property of CSS and GPS sources \citep{Odea98}. Also the linear projected jet size of $\sim$2 kpc, derived from the turnover frequency, is consistent with CSS sources \citep{Schulz15}. Its radio luminosity at 5~GHz is $\sim3.8\times10^{42}$ \ergs ($7.4\times10^{25}$ W Hz$^{-1}$, \citealp{Schulz15}), which lies at the lower end of the CSS/GPS luminosity distribution \citep{Odea98}, but within the typical range of jetted NLS1s \citep{Berton18a}.\par
The history of its optical classification has been somewhat troubled. The source was originally included in the NLS1 class by \citet{Oshlack01}, that estimated the black hole mass for the first time as 5.4$\times$10$^6$ M$_\odot$\footnote{Note that they used a different cosmology, with q$_0$=0.5 and H$_0$ = 100 \kms\ Mpc$^{-1}$.}. However, the spectrum they analyzed was derived from \citet{Drinkwater97}, and it unfortunately showed an issue in the y-axis of their Fig.~2, where the flux was underestimated by a factor 200 with respect to the original paper. This error propagated throughout their work and in the subsequent literature based on it. The low signal-to-noise ratio of the spectrum, furthermore, hampered a fully reliable classification of the source as an NLS1. Indeed, due to its seemingly weak Fe II emission, PKS 2004-447 was considered to be a possible narrow-line radio galaxy (NLRG) or a Type 2 AGN \citep{Zhou03, Sulentic03, Komossa06}, although \cite{Gallo06a} noted that the presence of strong Fe II multiplets is not always included in the definition of NLS1, and therefore preferred an NLS1 classification. More recently, a new estimate of the black hole mass from optical spectroscopy is 7$\times$10$^7$ M$_\odot$ \citep[see][arXiv:1409.3716v4, footnote 10]{Foschini15}, close to the value 9$\times$10$^7$ M$_\odot$ derived from the K-band bulge luminosity \citep{Kotilainen16}. A significantly higher value, 6$\times$10$^8$ M$_\odot$, was instead derived using optical spectropolarimetry \citep{Baldi16}. \par
PKS 2004-447 was eventually included in the NLS1 class due to its multiwavelength properties, which are similar to those of other $\gamma$-ray NLS1s. Its X-ray spectrum, in particular, is well described by a single power law, likely due to the non-thermal emission of the relativistic jet \citep{Kreikenbohm16}. A soft X-ray excess was detected in \emph{XMM-Newton} observations \citep{Gallo06a, Foschini09}, although its presence is not confirmed in all observations and is correlated to a high flux level \citep{Kreikenbohm16, Gokus21}. It seems to be the high-energy tail of the synchrotron emission \citep{Foschini09, Foschini20}. It is worth noting that, unlike what is seen in other $\gamma$-NLS1s \citep[e.g., see PMN J0948+0022,][]{Foschini12}, PKS 2004-447 has a less prominent variability both in X-rays and in radio \citep{Schulz15, Kreikenbohm16}. \par

\begin{table}[!t] 
    \centering
    \caption{Observational details for spectra.}
    \label{tab:data}
    \scalebox{0.9}{
    \begin{tabular}{l c c c c}
    \hline\hline
    Obs. date & Tel. & Inst. & Exp. time & $\lambda/\Delta\lambda$ \\
    \hline
    1984-05-02 & AAT & RGO-FORS/250B & ? & 150 \\ 
    2015-10-17 & VLT & FORS2/Gris300V & 4$\times$150 & 440 \\
    2015-10-22 & VLT & FORS2/Gris300V & 4$\times$150 & 440 \\
    2015-11-09 & VLT & FORS2/Gris300V & 4$\times$150 & 440 \\
    2016-03-18 & VLT & FORS2/Gris300V & 4$\times$150 & 440 \\
    2016-03-29 & VLT & FORS2/Gris300V & 4$\times$150 & 440 \\
    2016-04-02 & VLT & FORS2/Gris300V & 4$\times$150 & 440 \\ 
    2019-04-16 & ClT & LDSS3/VPH-ALL & 3$\times$300 & 860 \\
    2019-10-31 & NTT & EFOSC2/Grism\#5 & 3$\times$400 & 435 \\
    2019-11-29 & DuT & WFCCD/GrismB & 4$\times$1200 & 2350 \\
    \hline
    \end{tabular}
    }
    \tablefoot{Columns: (1) observation date; (2) Telescope. List of acronyms: Anglo-Australian Telescope (AAT), Very Large Telescope (VLT), Clay Telescope (ClT), New Technology Telescope (NTT), duPont Telescope (DuT); (3) Instrument and dispersion element used for observations. List of acronyms: Royal Greenwich Observatory spectrograph, Faint Object Red Spectrograph (RGO-FORS), Focal Reducer/low dispersion Spectrograph 2 (FORS2), Low Dispersion Survey Spectrograph 3 (LDSS3), ESO Faint Object Spectrograph and Camera 2 (EFOSC2), WFCCD (Wide-field CCD); (4) exposure time (in seconds); (5) spectral resolution. }
\end{table}

\section{Observations and data reduction}
The calibrated spectrum observed with the Anglo-Australian Telescope (AAT) on 1984-05-02 was extracted from the paper by \citet{Drinkwater97}. The spectra from the FOcal Reducer and low dispersion Spectrograph 2 (FORS2) mounted on the Very Large Telescope (VLT) were originally obtained to get a reliable optical classification of the source, and study its optical variability (program ESO/096.B-0256, P.I. Kreikenbohm). The data from the Las Campanas Observatory (Clay and Dupont telescopes) were collected to monitor the low state of the source and its $\gamma$-ray flare. Finally, the spectrum in the flaring state was observed in the framework of a larger NLS1 study carried out with the ESO Faint Object Spectrograph and Camera 2 (EFOSC2) at the New Technology Telescope (NTT, program ESO/0104.B-0587, P.I. Berton). The technical details of each spectrum are reported in Table~\ref{tab:data}. For all these new observations, we carried out the standard data reduction with IRAF, with bias and flat-field correction, and wavelength and flux calibration. \par
In all cases, we corrected the spectrum for Galactic absorption by using A(V)$=$0.091 \citep{Schlafly11} and assuming a reddening law with R$_v$ = 3.1 \citep{Cardelli89}. After this step we corrected the spectrum for redshift, z$=$0.240, and subtracted the AGN continuum by fitting it with a power law. We did not try to model the host galaxy contribution, because no absorption lines from the host are visible in the spectra. Furthermore, at the relatively high redshift of PKS 2004-447 the host contribution is expected to be negligible \citep{Letawe07}. \par
To account for the different observing facilities and conditions among our spectra, we decided to use the total flux of the [O III]$\lambda$5007 line as a reference to rescale the spectra. Since this line originates in the narrow-line region (NLR), which is significantly larger than the BLR and much farther away from the nucleus, its flux is expected to remain constant over several years \citep{Peterson04}. Since the VLT spectra are those with the highest signal-to-noise ratio (S/N), we decided to use their median [O III] flux as a reference, excluding the noisiest spectra. The first (2015-10-17) and the fifth observing nights (2016-03-29), indeed, were likely affected by some passing clouds, so they were not used. The reference flux we adopted is 4.84$\times$10$^{-15}$ \ergs cm$^{-2}$, based on fitting the line profile with a double Gaussian. The exact measurement procedures are described in Sect.~\ref{sec:o3} and in Sect.~\ref{sec:timevar}. 

\begin{figure}
    \centering
    \includegraphics[width=\hsize]{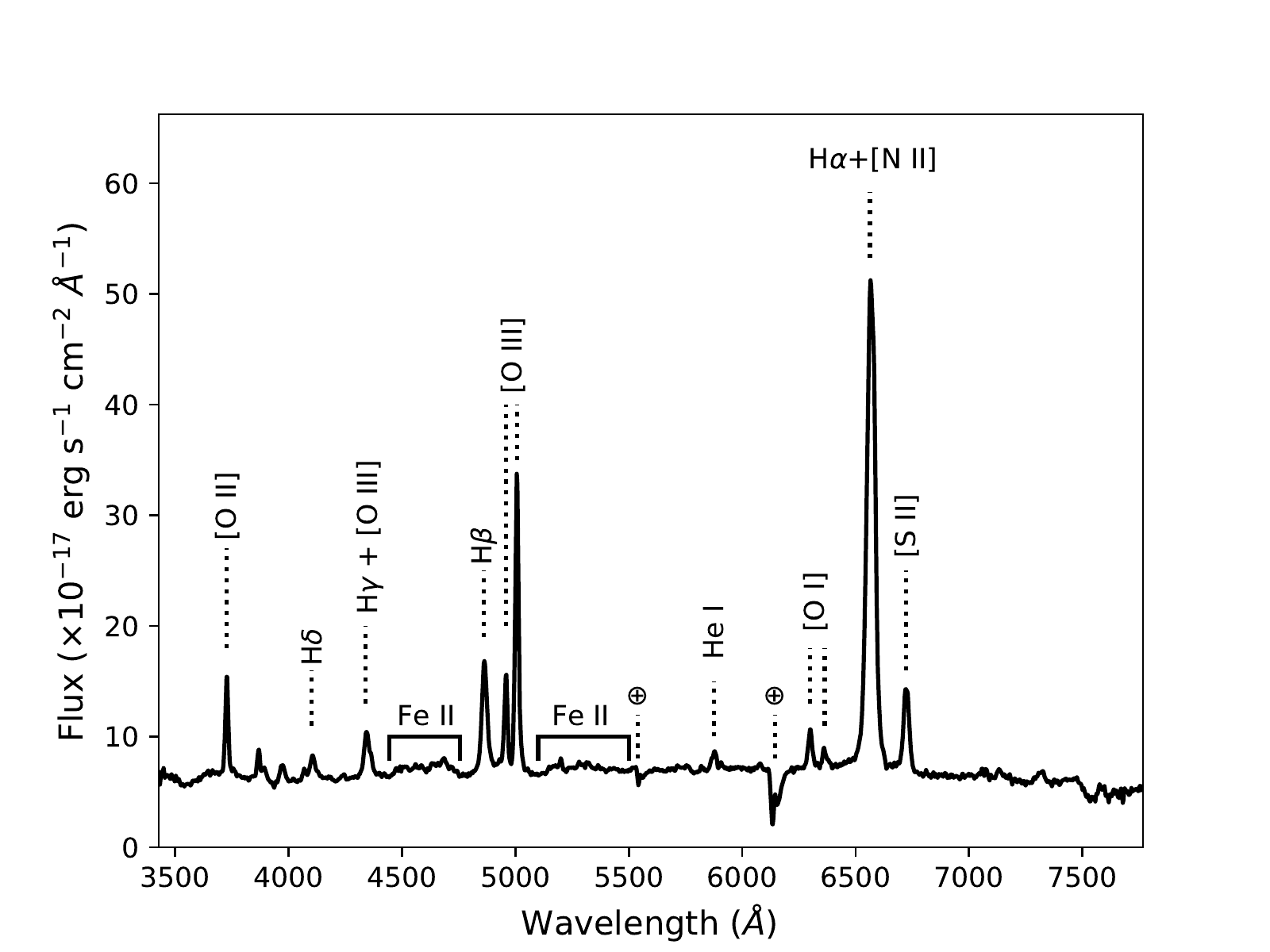}
    \caption{Redshift-corrected optical spectrum of PKS 2004-447, observed by FORS2. The most prominent emission lines are identified. The telluric absorptions are marked with the $\oplus$ symbol.}
    \label{fig:pks_fors}
\end{figure}

\section{Line profiles}
To carry out an analysis of the line profiles, it is crucial to have a spectrum with a high S/N. Indeed, this parameter is crucial to examine in detail the wings of profiles, and to obtain an accurate decomposition \citep[e.g.,][]{Jarvela20}. Therefore, we combined together all the spectra obtained by the VLT, reaching a S/N$\sim$90 in the continuum around 5100\AA. Although we may lose some information on the variability in the continuum and permitted lines over the six months of observations, this operation is necessary to ensure a good quality spectrum that can be studied in detail. The combined spectrum is shown in Fig.~\ref{fig:pks_fors}. All of the fitting procedures that follow have been performed with our own \texttt{Python} code \citep{Harris20}. An analysis of the variability will instead be carried out in Sect.~\ref{sec:timevar}. In our line profile analysis, we focused on the most prominent lines: H$\beta$, [O III]$\lambda\lambda$4959,5007, the [S II]$\lambda\lambda$6716, 6731 doublet, and the H$\alpha$+[N II]$\lambda\lambda$6548,6584 complex. \par

\subsection{Fe II multiplets}
\label{sect:feii}
\begin{figure}
    \centering
    \includegraphics[width=\hsize]{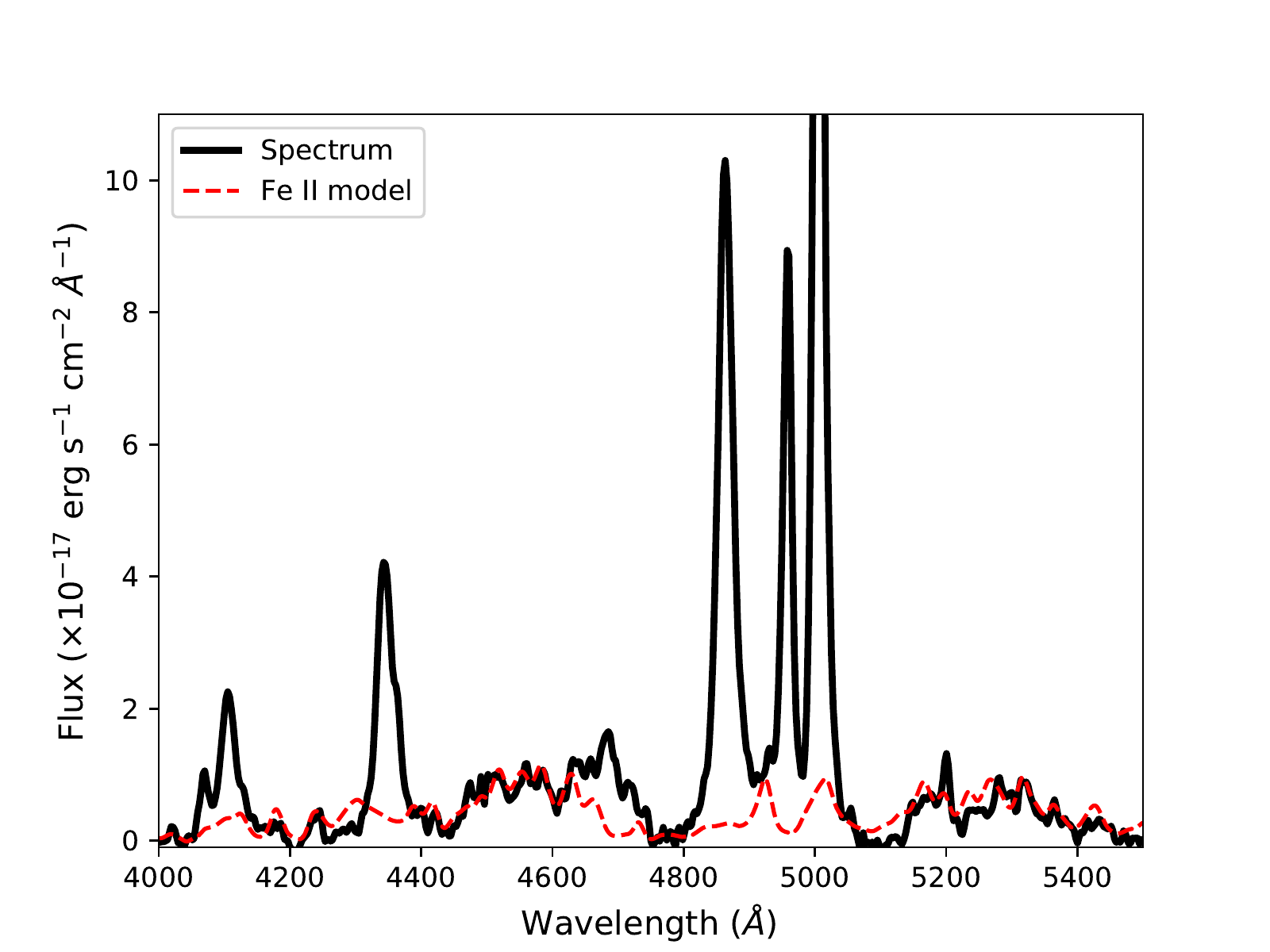}
    \caption{Fit of the Fe II multiplets in the H$\beta$ region. The original spectrum, redshift corrected and continuum subtracted, is represented by the solid black line. The Fe II multiplets from \citet{Marziani09} are the red dashed line. The spectrum was cut off for visual purposes.}
    \label{fig:fe_fit}
\end{figure}
The H$\beta$+[O III] region of NLS1s, between 4000 and 5500\AA, is typically characterized by the presence of Fe II multiplets. These lines lie very close to the other emission features, and they are often blended with them. In particular, they can affect significantly the red wings of the [O~III]$\lambda$5007 line and of the H$\beta$ line. To reproduce the multiplets, we used the templates based on photoionization models provided by \citet{Marziani09}, adopting a FWHM for the Fe II of $\sim$1500 \kms (comparable to that of H$\beta$ broad, see Sect.~\ref{sec:hb}) and rescaling it to match the observed spectrum. The best fit is shown in Fig.~\ref{fig:fe_fit}. The typical errors produced by Fe II subtraction were already estimated by \citet{Cracco16}, and they are of the order of 10\% of the flux. The template we used yields a flux of Fe II on the blue side of H$\beta$ of (3.1$\pm$0.3)$\times10^{-15}$ \ergs cm$^{-2}$. This value will be later used to calculate the R4570 parameter.


\subsection{[O III] lines}
\label{sec:o3}
\begin{figure}
    \centering
    \includegraphics[width=\hsize]{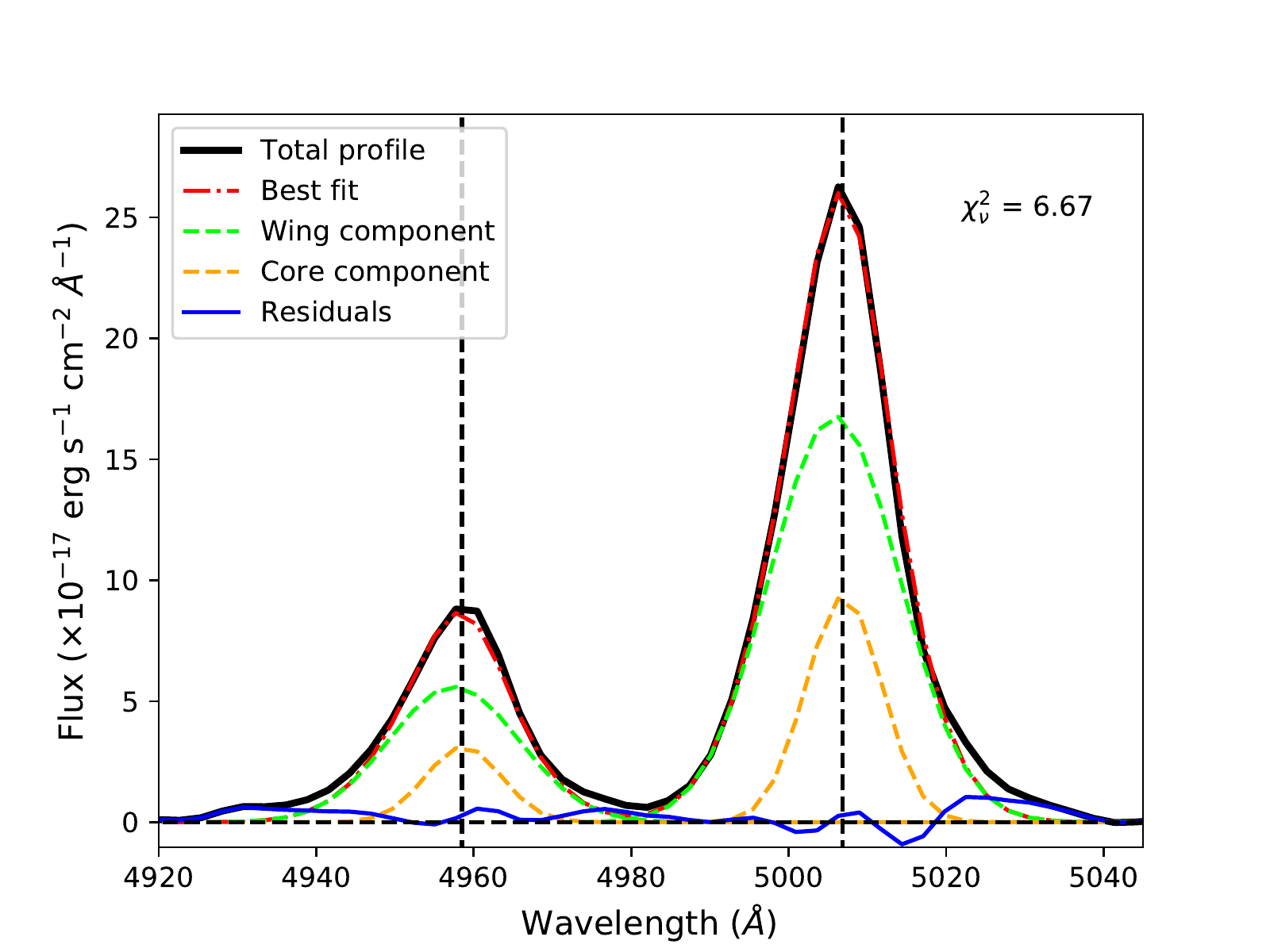}
    \caption{Fit of the [O~III] lines after continuum and Fe II subtraction. The spectrum is represented by the black solid line. The two Gaussian components, represent the line core (orange dashed line) and wing (green dashed line), respectively. The best fit is represented by the red dot-dashed line, the residuals by the solid blue line, and the horizontal dashed line represents the zero level of continuum. The restframe wavelength of the two lines is indicated by the vertical dashed lines. Due to the low spectral resolution, the core component is unresolved.}
    \label{fig:o3}
\end{figure}

After subtracting the Fe II multiplets, we modeled the [O III]$\lambda\lambda$4959, 5007 lines. Usually these lines show two separate components. The first one is a narrow core component, which is associated with the gas of the NLR and typically has the same redshift as the host galaxy. The second is a broad wing usually interpreted as a sign of outflowing gas. Therefore, the two lines should be modeled with four Gaussian functions. To reduce the number of free parameters, we introduced some constraints on the $\lambda$4959 line. The flux ratio between the components was fixed to the theoretical value of 1/3 \citep{Dimitrijevic07}, the FWHM of the core and the wing were forced to be the same in both lines, and the relative shift between the two components was also fixed to be the same in both lines. The measurements were performed by using a Monte Carlo method. We repeated the fit one thousand times while adding every time to the line profiles a different Gaussian noise proportional to the noise in the continuum. The latter is also used to estimate the $\chi^2_\nu$. We finally used the median value for each parameter and its standard deviation. The result of the fit is shown in Fig.~\ref{fig:o3}, where the \chired $\sim 6.7$ is also reported. The total flux we obtained for the two lines is 6.45$\times10^{-15}$ \ergs\ cm$^{-2}$, with one quarter of the flux in the $\lambda$4959 line and the rest in the $\lambda$5007 line. The fit indicates that the core component is unresolved. This result may not be real, but only a product of the limited spectral resolution. Therefore, no robust conclusion can be obtained on the [O~III] line components. \par
We also tried to reproduce the [O~III] lines by modeling them with a skewed Gaussian function. The function can be expressed as 
\begin{equation}
  f(\lambda) = \frac{A_0}{\sigma_s \sqrt{2\pi}} \exp{\left[-\frac{(\lambda - \lambda_0)^2}{2\sigma_s^2}\right]}  \left[ 1 + \mathrm{erf}\left( \frac{\alpha}{\sqrt{2}} \frac{(\lambda - \lambda_0)}{\sigma_s}\right)  \right] \; ,
\end{equation}
where $\lambda_0$ represents the central wavelength, $\sigma_s$ represents the width of the skewed Gaussian, $\alpha$ is the skewness parameter, and $A_0$ is a constant. Physically, this model seem to indicate that the ionized gas producing the [O III] lines is distributed in a bipolar outflow, with one of the outflows partially obscured by an intervening medium (i.e., the central engine). We fixed the parameters of the $\lambda$4959 line as described for the previous model. This new function provides a slightly worse representation of the line profile than the double Gaussian ($\chi^2_\nu=13.4$, $\Delta\chi^2_\nu=-6.7$). However, it is worth noting that the skewness parameter is $\alpha = -1.07\pm0.07$, suggesting a slight asymmetry on the blue side. This may indicate that the receding outflow could be partially obscured, as we would expect in a system observed not too far from its symmetry axis as a type 1 AGN. \par

\begin{figure*}[!ht]
    \centering
    \begin{minipage}[t]{.48\textwidth}
    \centering
	\includegraphics[trim={0cm 0cm 0cm 0cm}, width=\textwidth]{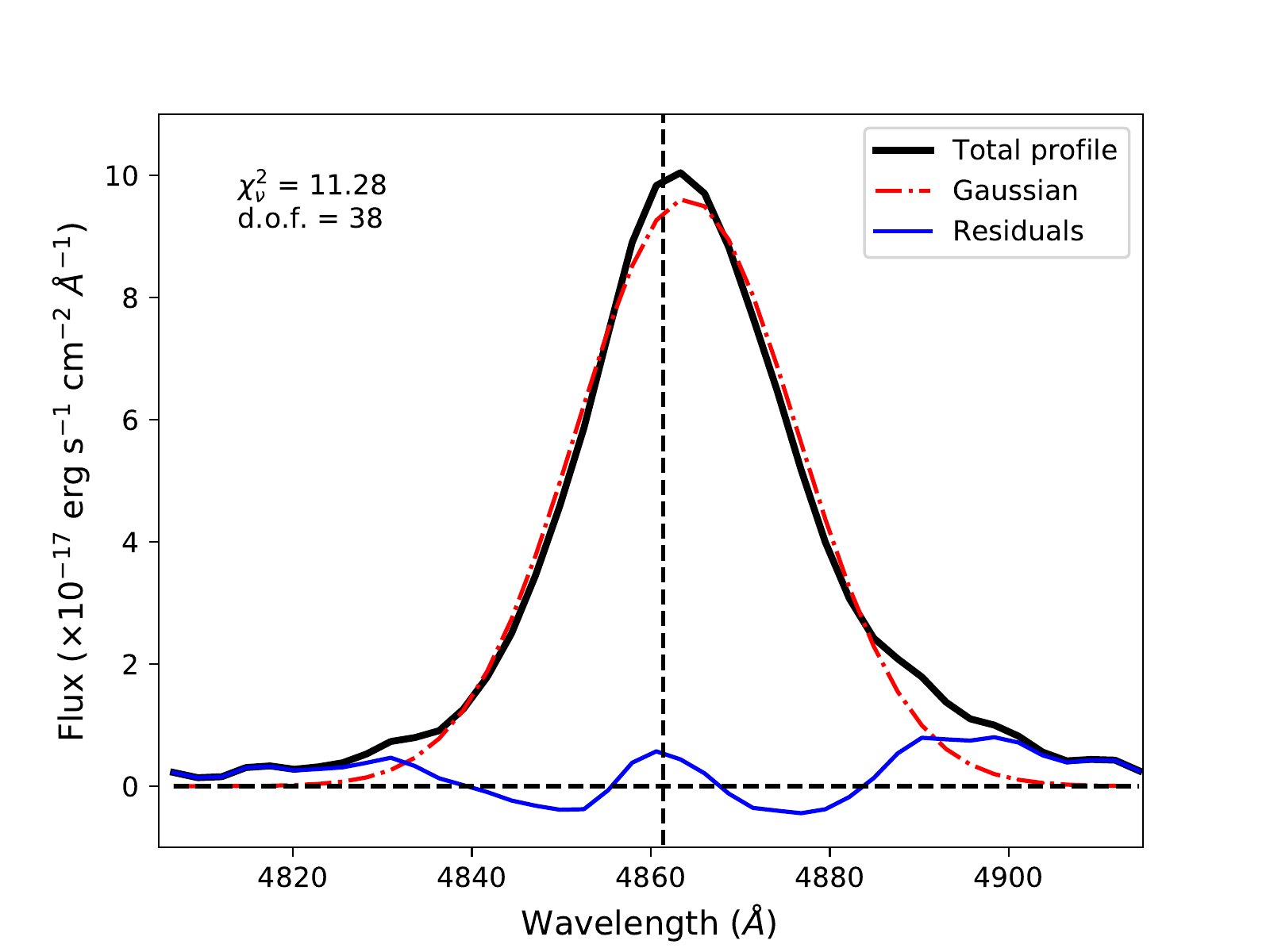}
    \end{minipage}
        \hfill
    \begin{minipage}[t]{.48\textwidth}
    \centering
	\includegraphics[trim={0cm 0cm 0cm 0cm}, width=\textwidth]{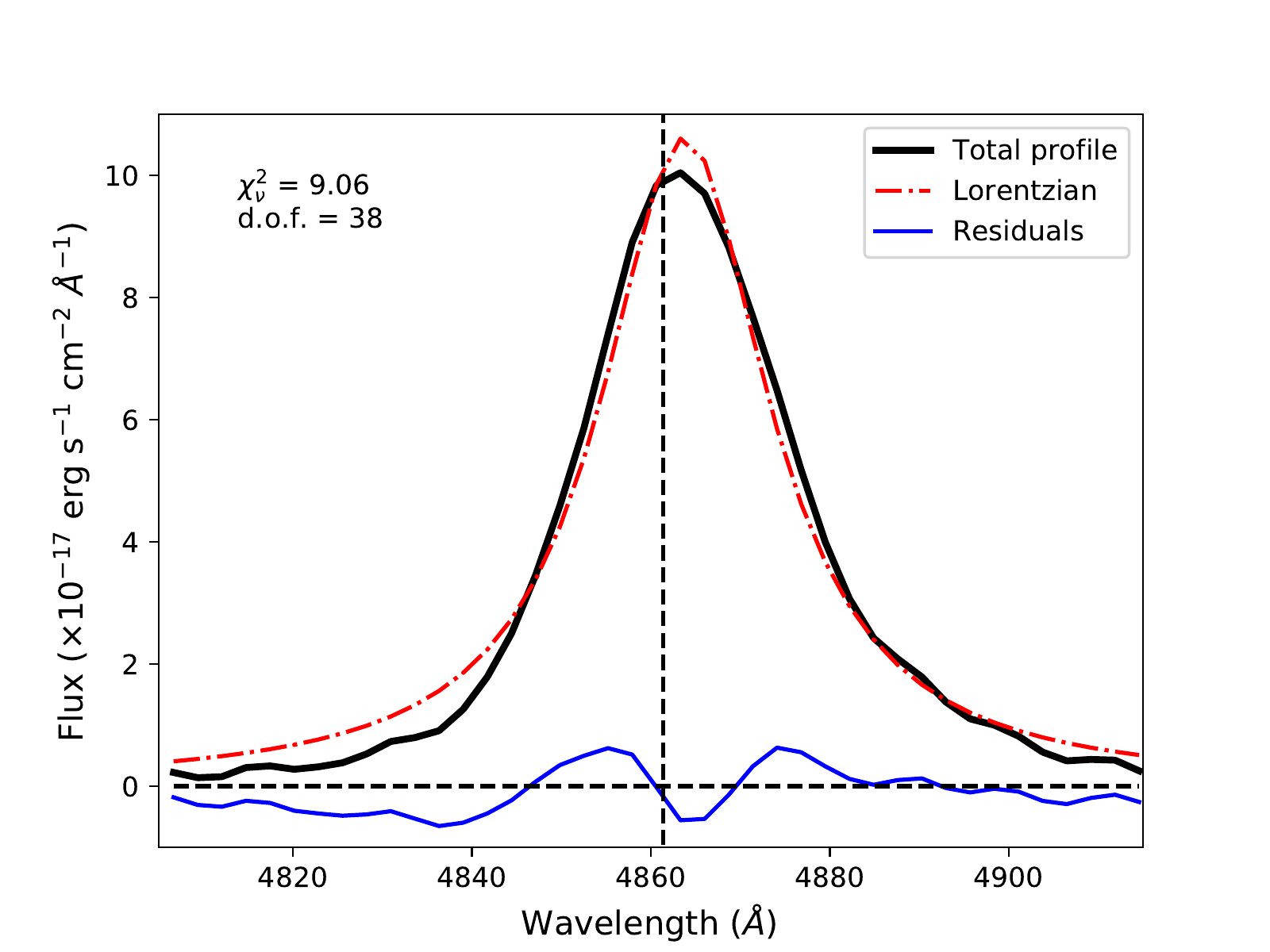}
    \end{minipage}     
            \hfill
    \begin{minipage}[t]{.48\textwidth}
    \centering
	\includegraphics[trim={0cm 0cm 0cm 0cm}, width=\textwidth]{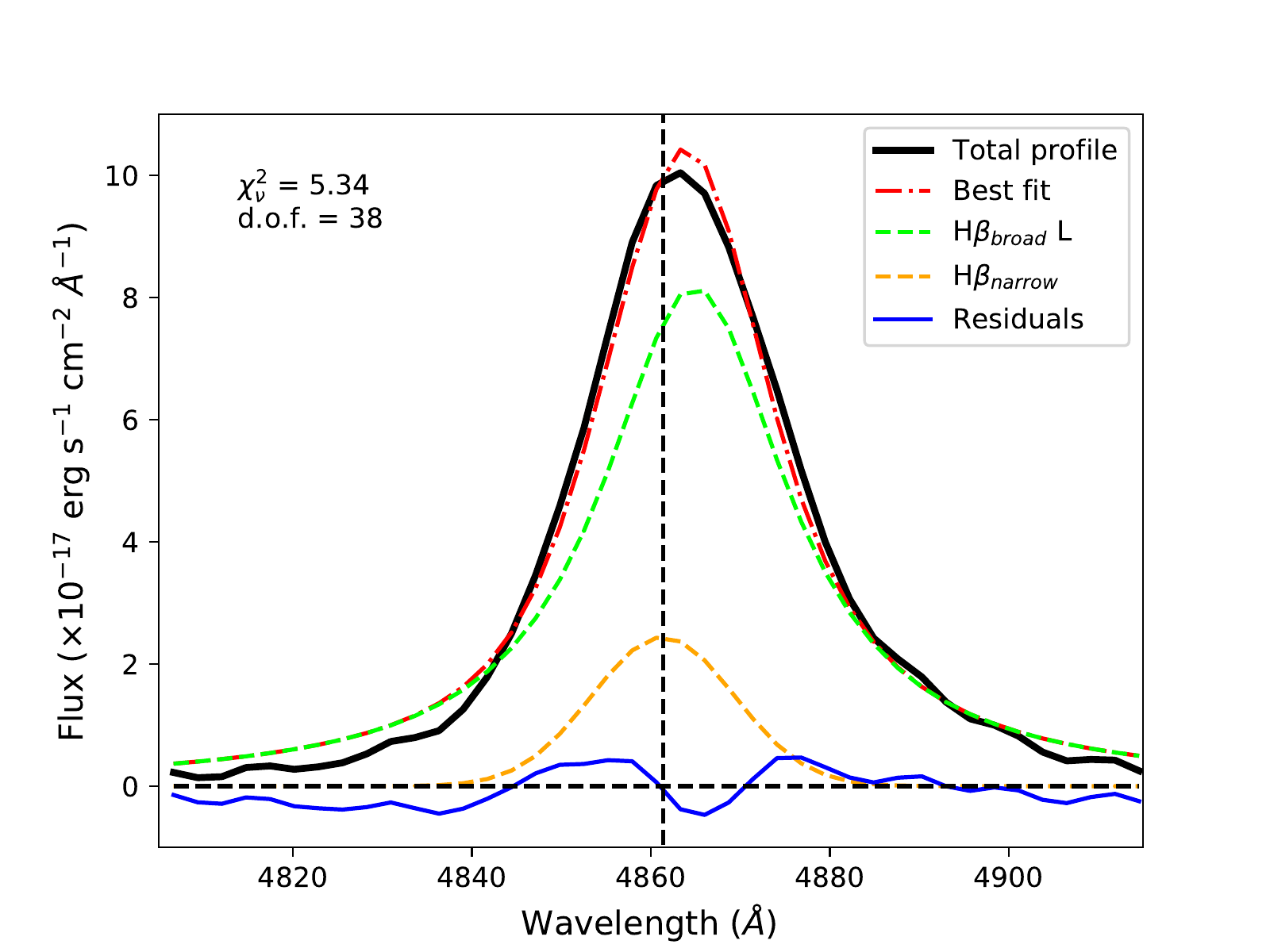}
    \end{minipage}   
            \hfill
    \begin{minipage}[t]{.48\textwidth}
    \centering
	\includegraphics[trim={0cm 0cm 0cm 0cm}, width=\textwidth]{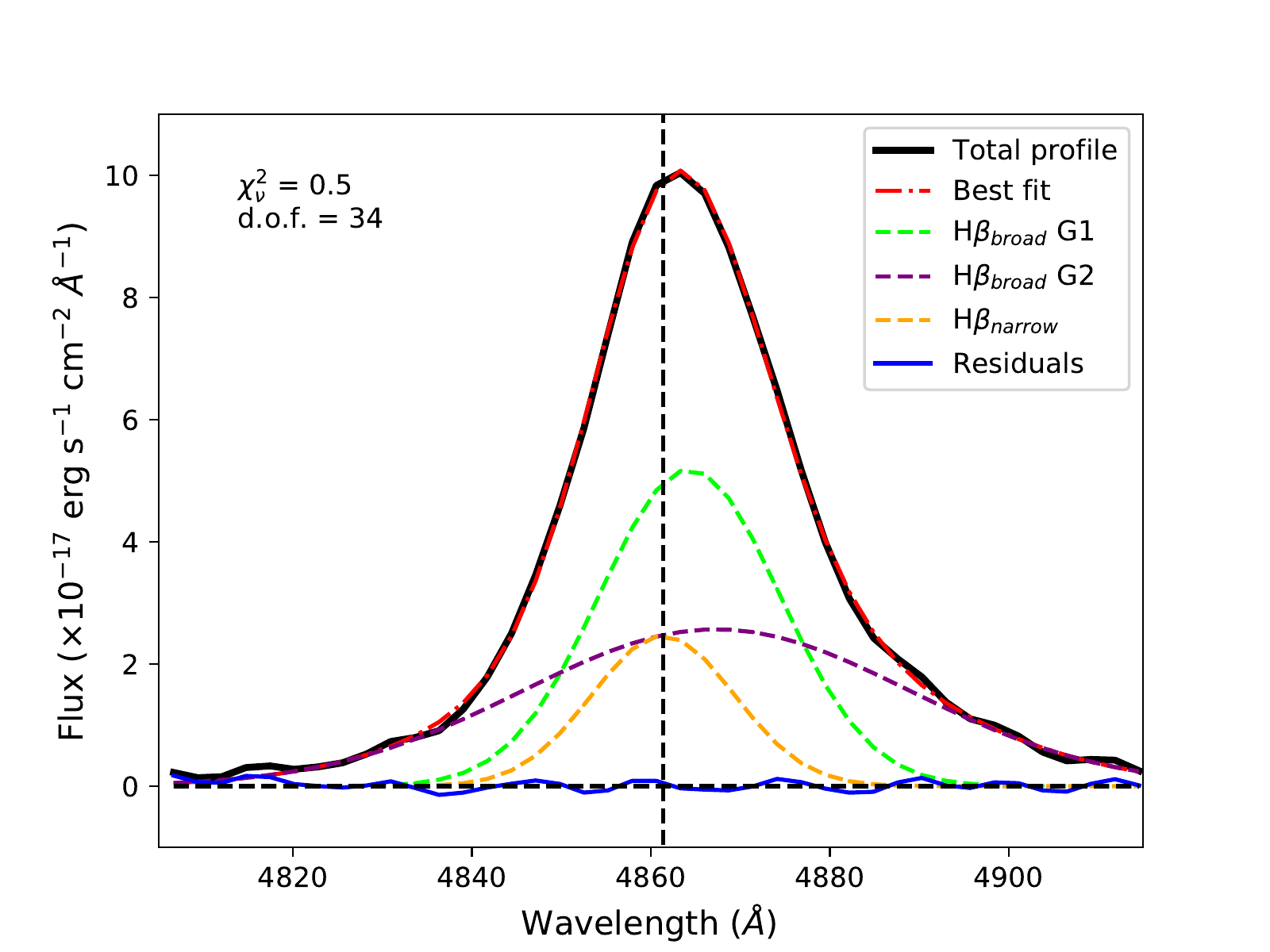}
    \end{minipage}   
    \caption{Fit of the H$\beta$ line with different functions. The original spectrum is the solid black line, the residuals of the model subtractions are the blue solid line, the horizontal black dashed line represents the zero level of continuum, while the vertical dashed line represent the restframe wavelength of H$\beta$. The values of \chired\ and degrees of freedom (d.o.f.) for each fit is shown in the top left corner of each figure. \textit{Top left:} fit with a single Gaussian function, represented by the red dot-dashed line. \textit{Top right:} fit with a single Lorentzian function, indicated by the red dot-dashed line. \textit{Bottom left:} fit with a Lorentzian function, indicated by the green dashed line, and a narrow Gaussian function, represented by the orange dashed line. The best fit is indicated by the red dot-dashed line. \textit{Bottom right:} fit with three Gaussian functions, indicated by the green, purple, and orange dashed lines. The narrow component is the orange function. The best fit is indicated by the red dot-dashed line.}
    \label{fig:profiles}
\end{figure*}

\begin{figure*}[!ht]
    \centering
    \begin{minipage}[t]{.48\textwidth}
    \centering
	\includegraphics[trim={0cm 0cm 0cm 0cm}, width=\textwidth]{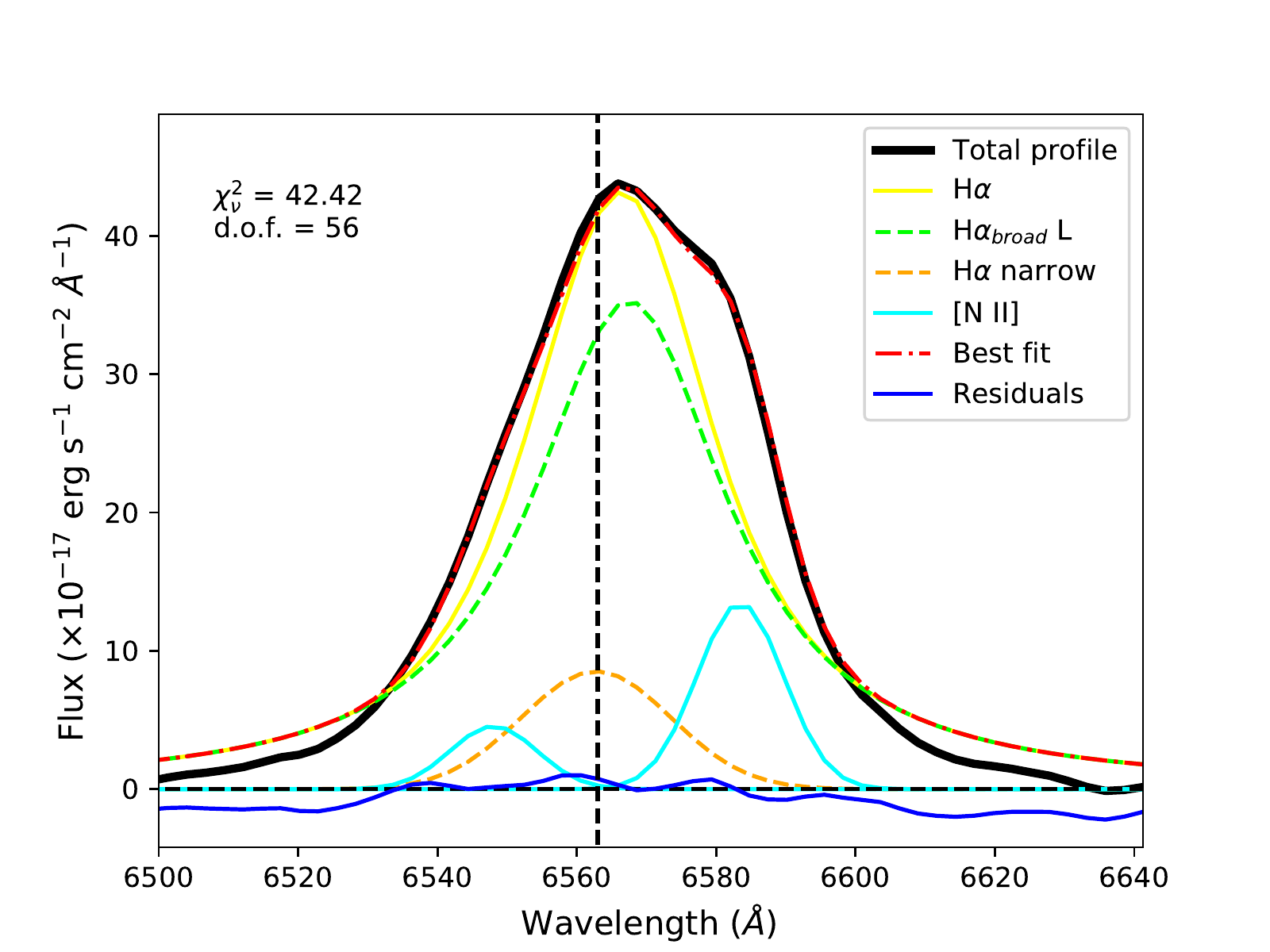}
    \end{minipage}
        \hfill
    \begin{minipage}[t]{.48\textwidth}
    \centering
	\includegraphics[trim={0cm 0cm 0cm 0cm}, width=\textwidth]{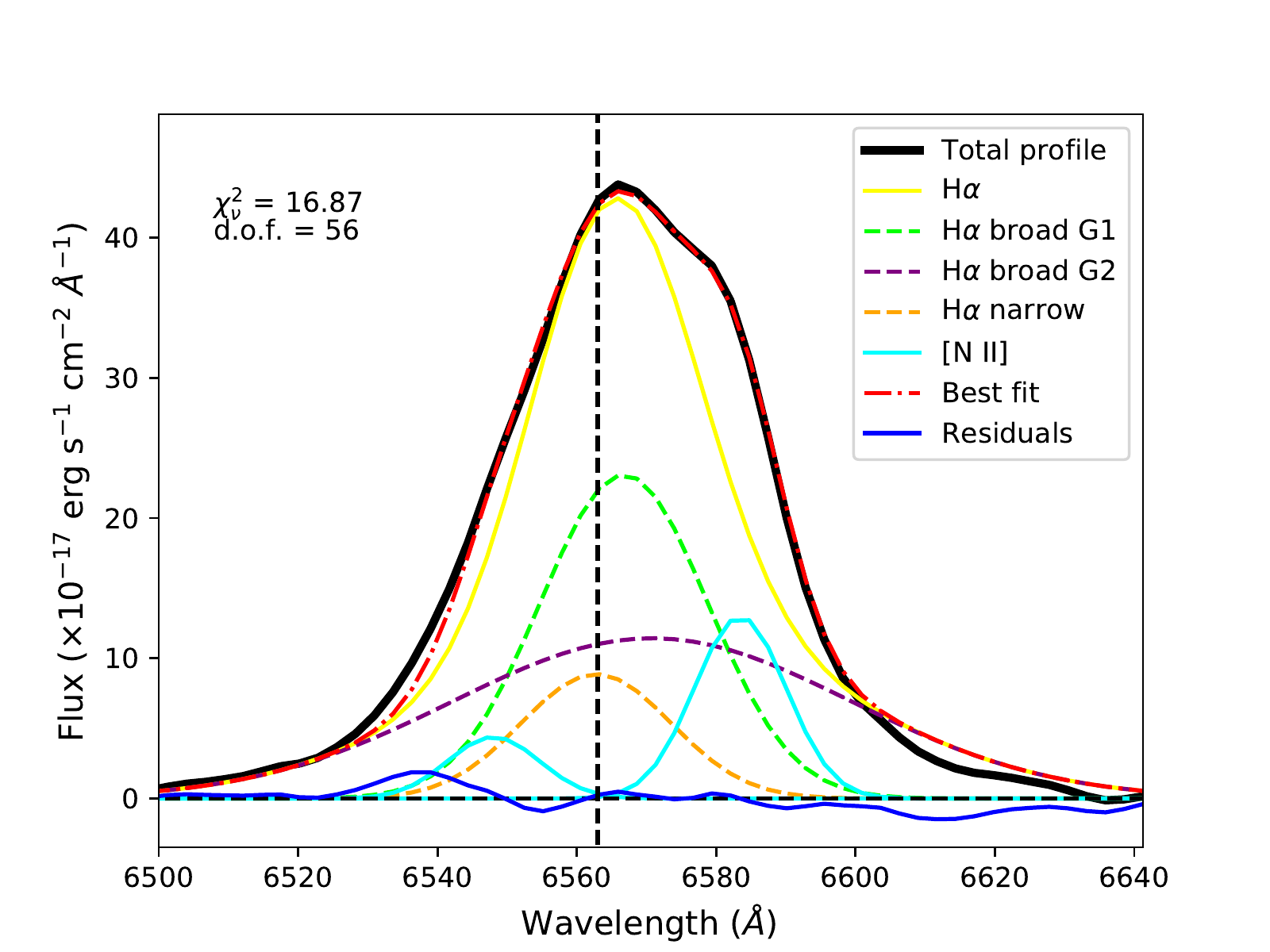}
    \end{minipage}    
    \caption{Fitting of the H$\alpha$+[N II] complex. The original spectrum is the solid black line, the residuals of the model subtractions are the blue solid line, the horizontal black dashed line represents the zero level of continuum, while the vertical dashed line represent the restframe wavelength of H$\alpha$. Two Gaussians reproduce the two [N II] lines, represented by the cyan solid line, and one Gaussian reproduces the H$\alpha$ narrow component, indicated by the orange dashed line. The yellow solid line is the actual H$\alpha$ profile, while the best fit is indicated by the red dot-dashed line. \textbf{Left panel:} the broad component of H$\alpha$ is reproduced with one Lorentzian function, reproduced by the green dashed line. \textbf{Right panel:} the broad component of H$\alpha$ is reproduced with two Gaussians, indicated by the green and purple dashed lines.}
    \label{fig:ha}
\end{figure*}

\subsection{H$\beta$ line}
\label{sec:hb}
The modeling of H$\beta$, shown in Fig.~\ref{fig:profiles}, was carried out after the subtraction of the Fe II multiplets. At first we attempted simple fits with a single Gaussian profile or a single Lorentzian profile. The single Gaussian profile is the worst fit (\chired = 11.28), because it cannot reproduce correctly neither the core nor the wings of the line. A single Lorentzian provides a slight improvement (\chired = 9.06), but it shows that the line profile is not perfectly symmetric, since the Lorentzian function fits the red side of the line well, but fails to reproduce the blue side and the peak. \par
Since neither function is good enough to fully reproduce the whole profile, we first added to the Lorentzian function a single Gaussian to represent the narrow component of H$\beta$ (LG model). Its flux was fixed at 1/10 of the total flux of the [O III]$\lambda$5007 line, that is 4.8$\times10^{-15}$ \ergs cm$^{-2}$, since this ratio is often observed in type 1 AGN \citep{VeronCetty01}. Its FWHM was instead fixed to that of the [O II]$\lambda$3727 line (14.4\AA\ $\sim$ 1158 \kms, 936 \kms\ when corrected for instrumental resolution), while its central wavelength was left free to vary. We did not use the FWHM of [O III] as reference because high-ionization lines are significantly more perturbed than low-ionization lines such as [O II] \citep{Komossa08}. The [O I]$\lambda$6300 line would also be a good indicator but, as seen in Fig.~\ref{fig:pks_fors}, it is weaker than [O II] and therefore less apt for this. Nevertheless, it is worth noting that its FWHM$\sim$830 \kms\ (corrected for instrumental resolution) is not very different from that of [O II]. The addition of the narrow component leads to a significant statistical improvement of the fit (\chired = 5.34). However, the blue side of H$\beta$ is still not perfectly reproduced, and the same is true for the peak of the line. We therefore tried to reproduce the line using three Gaussians (3G model), two representing the broad component and one the narrow component. The latter had the same characteristics as before, while all the parameters of the two broad Gaussians were left free to vary. Statistically this produces the best fit for the line (\chired = 0.5), since it can reproduce the asymmetry of the profile by shifting the center of the very broad component. It is worth noting that such shift is consistent with the outflowing component that may be present in [O~III]. Using this model, we can derive the fundamental parameters of H$\beta$, that is FWHM(H$\beta)= 1617\pm8$ \kms, and total integrated flux F $= (3.17\pm0.01)\times10^{-15}$ \ergs\ cm$^{-2}$. \par
Using the LG model, we find that the width of the H$\beta$ broad component is FWHM(H$\beta_b) = 1577\pm10$ \kms, while the 3G model provides a FWHM(H$\beta_b) = 1804\pm62$ \kms, while the second-order moment of the line, defined as
\begin{equation}
    \sigma^2 = \frac{\int \lambda^2 F(\lambda) d\lambda}{\int F(\lambda)d\lambda} - \left(\frac{\int \lambda F(\lambda) d\lambda}{\int F(\lambda)d\lambda} \right)^2 \; ,
    \label{eq:sigma}
\end{equation}
is $\sigma$(H$\beta_b) = 1007\pm20$ \kms. The $\sigma$ of the LG model cannot be estimated because the second-order moment of a Lorentzian function is, by construction, infinite. The total flux of the line broad component estimated by the LG model is $(2.76\pm0.01)\times10^{-15}$ \ergs cm$^{-2}$, while the 3G model provides $(2.67\pm0.02)\times10^{-15}$ \ergs cm$^{-2}$. Using the mean of these two values, we calculated R4570 = F (Fe II) / F (H$\beta_b$) = 1.14$\pm$0.14. According to the classification by \citet{Marziani18b}, PKS 2004-477 therefore belongs to population A3 of the quasar MS.

\subsection{H$\alpha$ region}
We modeled the H$\alpha$ profile by fixing it to the H$\beta$ parameters. Specifically, we fixed the flux ratio of the different components in the LG and 3G model, the velocity shifts between the components, and the velocity associated to the FWHM of each component. This essentially leaves only one free parameter to be determined, that is the height of the narrow component. Finally, we added two more Gaussians to reproduce the [N II]$\lambda\lambda$6548,6584 lines, which are blended with H$\alpha$ due to its width and to the low resolution. Their FWHM is identical in both lines but free to vary, their flux ratio was fixed to the theoretical value of 2.95, and their positions were fixed to restframe. Therefore this leaves two free parameters. \par
This lead to the results shown in Fig.~\ref{fig:ha}. Both fits are not perfect, because the H$\beta$ profile used as reference seems to have a more prominent red wing, which is not observed in H$\alpha$. Anyway, the adoption of the double Gaussian instead of the Lorentzian to reproduce the broad component leads to a very significant improvement in the \chired, with $\Delta$ \chired = 25.55 (\chired\ values are reported in Fig.~\ref{fig:ha}). This seems to indicate that the double Gaussian is better at reproducing the broad component of both H$\alpha$ and H$\beta$. The reason for the asymmetry observed in H$\beta$ but not in H$\alpha$ could be some residual Fe II in the former that the template could not account for, or a real physical difference due to a different kinematics of the emitting gas. Only better data will allow us to disentangle between these two possibilities. \par
After fitting H$\alpha$, we estimated the internal reddening due to the dust by studying the Balmer decrement. We calculated the ratio $\mathcal{R}$ between the flux of the narrow component of H$\alpha$ and H$\beta$, which is $\mathcal{R} \sim 4.86$. Assuming a theoretical ratio of 2.86, and following \citet{Cardelli89}
\begin{equation}
    A(V) = 7.215 \log\left(\frac{\mathcal{R}}{2.86}\right) \; ,
\end{equation}
we found an internal extinction A(V) = 1.66 mag. This result is in agreement with what was found by \citep{Gallo06a}, who derived A(V) = $1.9\pm1.5$ from the AAT spectrum\footnote{Since this estimate involves a line ratio, the issue with the y-axis in \citet{Oshlack01} is not important in this case, as it affects all lines with a multiplicative factor.}. As they already pointed out, this extinction is significantly higher than what can be estimated from the X-ray spectra, which instead show negligible absorption, as confirmed by more recent observations \citep{Kreikenbohm16, Berton19b, Gokus21}. They also suggested that a possible explanation for this is a very different gas/dust ratio than what is seen in the Milky Way, and the jet may play a role in this by transferring material from the nucleus into the NLR. However, given the significant amount of uncertainty on the A(V) value, we decided not to apply an additional correction for internal exctinction in the calculations that will follow. 

\subsection{Electron density and temperature}
\label{sec:temden}
To derive the physical parameters of the NLR, we reproduced the [S II]$\lambda\lambda$ 6716, 6731 doublet, in order to measure the electron density. The lines are only partially resolved, therefore we fitted them with two Gaussians with the same width as [O II] (in velocity), with fluxes and positions free to vary. The flux ratio we obtained is F(6716)/F(6731) $\sim$ 1.03. \par 
Furthermore, we measured the flux of the [O III]$\lambda$4363 line, that is detected blended with H$\gamma$ in our spectrum. Given that both lines are rather faint, we adopted a simple model to fit them, using one Gaussian function for H$\gamma$ and one for [O III]. This yields a flux of 0.29$\times10^{-15}$ \ergs\ cm$^{-2}$, and a ratio [F(4959) + F(5007)]/F(4363)$\sim$23.0. \par
Using these ratios, we can provide an estimate of the electron density and temperature. We used the \texttt{temden} task of IRAF, which is based on a 5-level atomic model described by \citet{Derobertis87}. The observed line ratios are reproduced if the electron density is $n_e = 7.6\times10^2$ cm$^{-3}$, and the electron temperature is $T_e = 3.1\times10^4$ K. While the density value we derived is rather typical for AGN \citep{Congiu17b}, it is worth noting that, if some internal absorption is present as possibly suggested in the previous section, the temperature value we obtained should be treated as a lower limit.

\section{Black hole mass}

To estimate the black hole mass, we used different techniques. All of them are based on the assumption that the gas orbiting the black hole is virialized. In this case, the black hole mass can be calculated using the virial theorem, 
\begin{equation}
M_{BH} = f\frac{R_{BLR} v^2}{G} \; ,
\label{eq:virial}
\end{equation}
where R$_{BLR}$ is the radius of the BLR, $v$ is the rotational velocity of the gas, $G$ is the gravitational constant, and finally $f$ is the so-called scaling factor. The weight of this factor is still largely unknown, therefore we will for now fix $f=1$, and later discuss the results under different assumptions. \par

\subsection{Dependence on line width and velocity}
The two key parameters to determine are the radius of the BLR and the rotational velocity. To obtain the former, we used two relations calibrated in the literature. Both of them are based on the assumption of photoionization equilibrium. The accretion disk radiation is what causes the formation of BLR lines and pushes away the clouds due to radiation pressure. If the ionizing continuum coming from the disk is strong, the BLR will have a large radius. Indeed, there is a relation which seems to connect the BLR radius measured via reverberation mapping technique and the luminosity of the continuum at $\lambda$5100\AA. The coefficients were estimated by \citet{Bentz13}, 
\begin{equation}
    \log\left(\frac{R_{BLR}}{\mathrm{l.d.}} \right) = (1.53\pm0.03) + (0.53\pm0.03) \log \left(\frac{\lambda L_\lambda(5100\AA)}{10^{44}\mathrm{erg\;s}^{-1}} \right)
    \label{eq:bentz} \; ,
\end{equation}
where the BLR radius is expressed in light days. However, as pointed out before, the ionizing continuum produced by the accretion disk is also responsible for the formation of the emission lines. Therefore, the intensity of the lines is also proportional to the accretion disk luminosity, and the BLR radius depends on the luminosity. The relation between these quantities was derived by \citet{Greene10}, 
\begin{equation}
    \log\left(\frac{R_{BLR}}{\mathrm{l.d.}} \right) = (1.85\pm0.05) + (0.53\pm0.04)\log \left(\frac{ L(H\beta)}{10^{43}\mathrm{erg\;s}^{-1}} \right) \; ,
    \label{eq:greene} 
\end{equation}
where L(H$\beta$) is the integrated luminosity of the line. In principle, this second method should be less contaminated by the jet contribution, which is a non-negligible factor in the $\lambda$5100 luminosity \citep{Berton15a}. \par
To derive the gas velocity, we used the H$\beta$ line decomposition previously described. The best results were obtained when the broad component was fitted by either two Gaussians or a Lorentzian function. For both of the profiles, we adopted the FWHM of the broad component as a proxy of the rotational velocity of the gas. In the case of the double Gaussian, we also estimated the second-order moment $\sigma$ of the broad component, defined in \ref{eq:sigma}. The use of $\sigma$ instead of FWHM provides generally better results, especially in low-contrast lines, and it is less affected by inclination effects and BLR geometry \citep{Peterson04, Peterson11, Peterson18}. We did not use this method for the Lorentzian profile because the $\sigma$ of a Lorentzian function is, by definition, infinite. \par
In conclusion, we had six different combinations to use for the calculation of this virial product, whose results are shown in Table~\ref{tab:masses}. The estimate of the errors on the mass was performed in two ways. The first source of error is in the fitting procedure, and we evaluated it using a Monte Carlo technique. We calculated the virial product one thousand times by varying the H$\beta$ profile adding a Gaussian noise proportional to the root mean square measured in the $\lambda$5100 continuum. This source of error is relatively small, typically around 0.03 dex, likely because of the high S/N of our spectrum. Another source of errors are the uncertainties on equations \ref{eq:bentz} and \ref{eq:greene}. We estimated one thousand times the BLR radius in both ways by applying to the coefficients a Gaussian noise proportional to their errors, and used these different values in the final calculation. Finally, we applied the normal propagation of errors to the virial theorem of equation \ref{eq:virial}. We calculated the weighted average of all the estimates obtained in Table~\ref{tab:masses}, and the resulting product is (6.0$\pm$0.4)$\times 10^6$ M$_\odot$. In conclusion, Table~\ref{tab:masses} shows that different methods can lead to statistically significant differences in the black hole mass calculation. Given the high S/N of the spectrum we used for the fitting procedure, the main source of error is not the fit itself, but instead the propagation of the errors on all the uncertain quantities such as the BLR radius. \par

\begin{table}[]
\caption{Product Rv$^2$/G calculated with different techniques.}
\label{tab:masses}
    \centering
    \begin{tabular}{c c c c}
    \hline
    Function & R$_{BLR}$ & v$_{rot}$ & log(Rv$^2$/G) \\
    \hline\hline
    3G & L(5100) & $\sigma$(H$\beta$) & 6.70$\pm$0.04$\pm$0.09 \\
    3G & L(5100) & FWHM(H$\beta$) & 7.20$\pm$0.05$\pm$0.33 \\
    3G & L(H$\beta$) & $\sigma$(H$\beta$) & 6.48$\pm$0.02$\pm$0.10 \\
    3G & L(H$\beta$) & FWHM(H$\beta$) & 7.00$\pm$0.03$\pm$0.37 \\
    LG & L(5100) & FWHM(H$\beta$) & 7.13$\pm$0.04$\pm$0.18 \\
    LG & L(H$\beta$) & FWHM(H$\beta$) & 6.93$\pm$0.01$\pm$0.21 \\
    \hline
    \end{tabular}
\tablefoot{Columns: (1) function used to reproduce the H$\beta$ line. 3G stands for three Gaussians (two for the broad component), and LG stands for Lorentzian and Gaussian (the Lorentzian function represents the broad component); (2) technique used to calculate the BLR radius, either L(5100\AA) by means of equation \ref{eq:bentz}, or L(H$\beta$) following equation \ref{eq:greene}; (3) proxy of the rotational velocity, either FWHM(H$\beta$) or $\sigma$(H$\beta$) calculated as in equation \ref{eq:sigma}; (4) logarithm of the virial product, that is the black hole mass assuming scaling factor f$=$1. The first error is due to the fitting procedure, while the second is the statistical error.}
\end{table}

\subsection{Dependence on the $f$ factor}
To this point we neglected the scaling factor $f$ which appears in Equation~\ref{eq:virial}, which is another major source of uncertainty in the black hole mass estimate. The scaling factor accounts for the difference between the mass obtained with the product $R_{BLR}v^2/G$ and the actual black hole mass. This difference strongly depends on the geometry and inclination of the BLR. If the BLR has a flattened geometry, it is clear that when observed pole-on there would be no velocity component parallel to the line of sight, thus causing a severe underestimate of the gas rotational velocity and, as a consequence, of the black hole mass \citep[e.g.,][]{Decarli08}. In case of a more sphere-like geometry this effect would be less evident. Our knowledge of the structure of the BLR is still relatively limited. While it is clear that Keplerian motion of the clouds is present \citep{Peterson99, Gravity18}, there may be some additional components such as turbulent vertical motion, possibly originating in a disk wind, that can significantly affect the value of $f$ \citep{Gaskell09, Kollatschny13}. A perfectly thin BLR is indeed clearly unphysical. The BLR clouds must be illuminated by the central engine ionizing photons. If the clouds were distributed on a thin disk, the outer clouds would not be reached by the disk radiation \citep{Collin06}. \citet{Shen14} suggested that $f$ is inversely proportional to the FWHM(H$\beta$), with larger values, up to $\sim$100, corresponding to small FWHMs. However, lower FWHM(H$\beta$) typically correspond to different line profiles, since type 1 AGN with narrow lines (e.g., NLS1s) tend to show Lorentzian profiles instead of Gaussian ones \citep{Sulentic00, Marziani01, Komossa08a, Cracco16}. Since Lorentzian profiles are possibly associated with the presence of a significant vertical structure in the BLR \citep{Kollatschny11, Kollatschny13a}, the effect of inclination may not be so prominent in NLS1s and population A objects \citep{Vietri18, Berton20a}. \par
Several estimates of the $f$ factor, both $f_\sigma$ and $f_{\rm FWHM}$, exist in the literature, and they are mostly based on reverberation mapping observations. Typical values can span between $\sim$0.8-5.0 \citep{Mandal21}. In the case of PKS 2004-447 this means that its black hole mass, using the weighted average estimated above, ranges between (4.8-30.0)$\times10^6$ M$_\odot$, within the typical range of NLS1s \citep{Peterson11}. A weak dependence on the FWHM(H$\beta$), and also on the ratio between the FWHM of the line and its dispersion, seem to be present. \citet{Collin06} calculated different values of $f$ depending on how the rotational velocity is estimated (FWHM, $\sigma$), on the ratio between FWHM and $\sigma$, and on the FWHM(H$\beta$) itself. In the case of PKS 2004-447, we decided to use their values, that are $f_\sigma=3.93$ for $\sigma$-based measurements, and $f_{\rm FWHM}=2.12$ for FWHM-based measurements. By applying these values to the virial products calculated in Table~\ref{tab:masses}, and taking the weighted average, we obtained a mass value of (1.5$\pm$0.2)$\times$10$^7$ M$_\odot$. In the following, we will use this value for our calculations. \par

\section{Time variability}
\label{sec:timevar}
\subsection{Lines and continuum}
\label{sec:lines}
\begin{figure}[!t]
    \centering
    \includegraphics[width=\hsize]{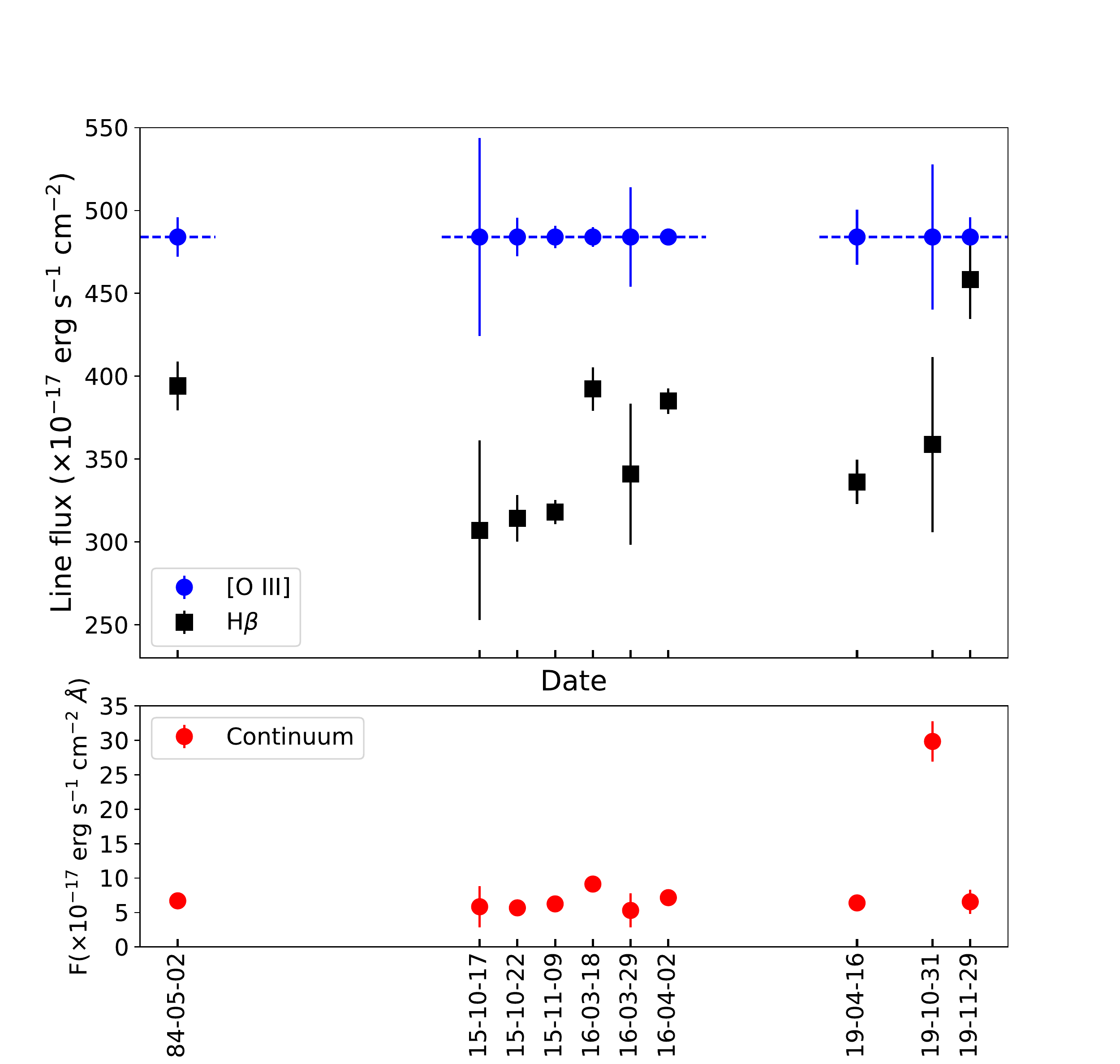}
    \caption{The variability measured in the continuum flux density and in the H$\beta$ line flux. The blue circles represent the [O III]$\lambda$5007 flux, which was forced to be the same in all spectra. The black squares represent the flux of H$\beta$, while the red circles in the bottom panel represent the continuum flux. }
    \label{fig:variability}
\end{figure}
Our goal is to study the long-term variability of PKS 2004-447 spectra. As previously mentioned, we accounted for the variability induced by different observing conditions by rescaling our spectra using the [O III]$\lambda$5007 as a reference. The measurement of the [O III] flux density was carried out after redshift correction, continuum subtraction, and Fe II multiplets modeling and subtraction. The exceptions to this are the spectra in which, because of the low S/N, we could not model the Fe II multiplets. They are 1984-05-02, 2015-10-17, 2016-03-29, and 2019-10-31. All the spectra were rescaled to the value derived in Sect.~\ref{sec:o3}, that is 4.84$\times 10^{-15}$ \ergs\ cm$^{-2}$. The two parameters we measured are the continuum flux at 5100\AA, and the H$\beta$ total flux. The errors on the line fluxes were estimated as previously described, while the uncertainty on the continuum at 1 $\sigma$ confidence level is provided by the standard deviation of a spectral region dominated by continuum emission with no (or very little) contamination by strong emission lines between 5075 and 5125 \AA. The results are shown in Fig.~\ref{fig:variability}. \par
\begin{figure}[!t]
    \centering
    \includegraphics[width=\hsize]{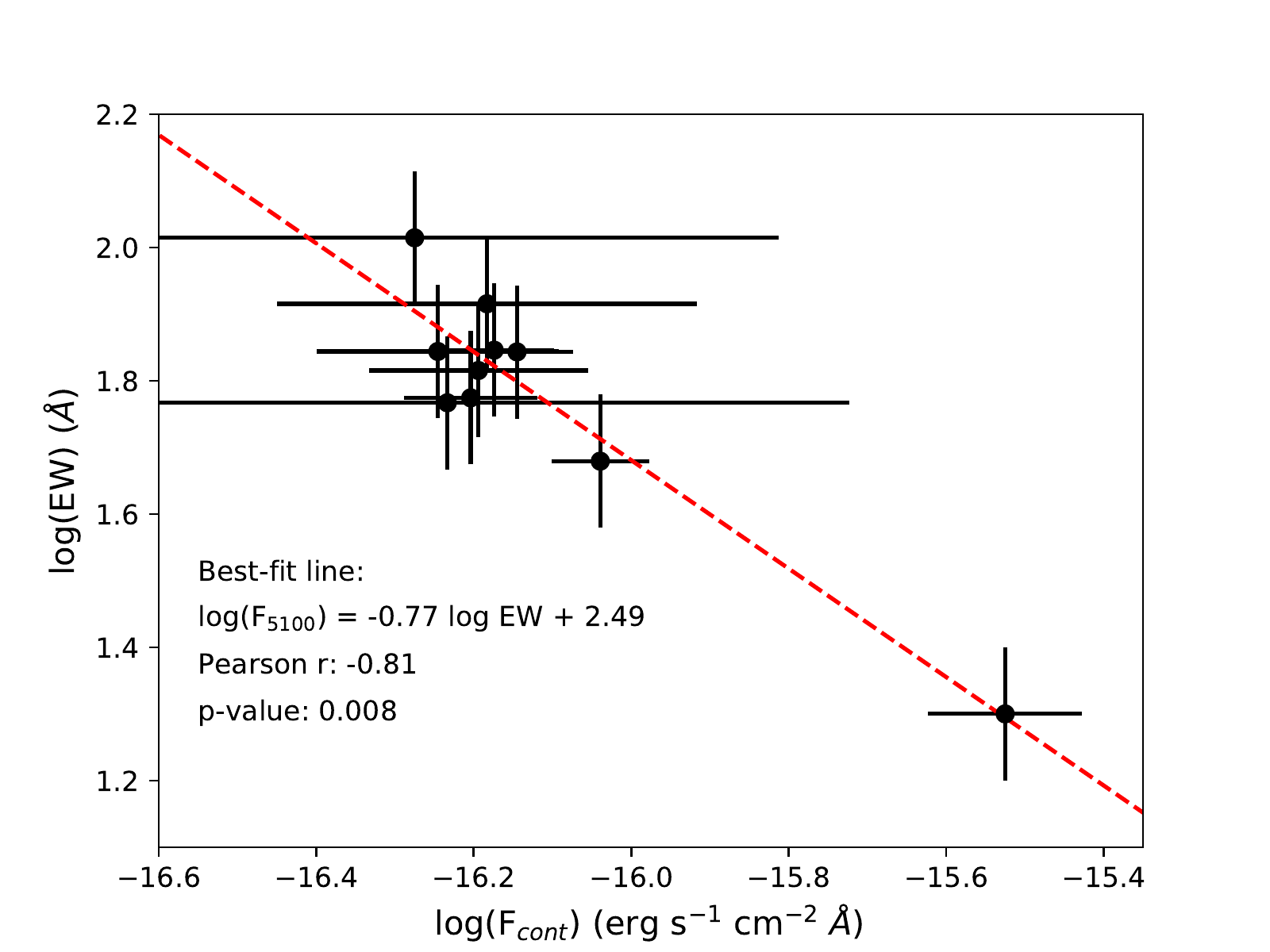}
    \caption{Anticorrelation between continuum flux at 5100\AA\ and equivalent width of the H$\beta$ emission line. The red dashed line indicates the best linear fit.}
    \label{fig:eqw}
\end{figure}
The most apparent result is the major flux increase that was observed in the continuum on 2019-10-31, which is remarkably followed by the highest measured H$\beta$ flux one month later (2019-11-29). This is due to the flaring activity of PKS 2004-447 measured in those days. The flare was detected in $\gamma$-rays by the \textit{Fermi} Satellite on 2019-10-25 \citep{Gokus19} and by AGILE until 2019-10-27 \citep{Verrecchia19}, with enhanced activity measured at other frequencies as well \citep{Dammando19, Berton19c, Blaufuss19}. Even if PKS 2004-447 has been included in the $\gamma$-ray catalogs since its very first detection soon after the launch of \textit{Fermi} \citep{Abdo09c}, this was its first recorded major flare \citep{Gokus19}. Indeed, our long-term spectroscopy shows that the continuum flux has been remarkably constant during the last forty years. The values we measured at each epoch are reported in Table~\ref{tab:variability}. Beside the flare, the continuum has varied between minimum and maximum values (5.3$\pm$2.4) and (9.1$\pm$0.6)$\times10^{-17}$ \ergs cm$^{-2}$ \AA$^{-1}$, recorded on 2016-03-29 and 2016-03-18, respectively. The H$\beta$ flux is instead more variable. A local maximum is seen in the 1984-05-02 spectrum. The H$\beta$ flux seems to also be increased in the three spectra following the maximum continuum value measured on 2016-03-18. Furthermore, we investigated the presence of a correlation between the continuum flux and the equivalent width of H$\beta$ (which we define as positive in emission lines). The result is shown in Fig.~\ref{fig:eqw}. A strong anticorrelation was found, with Pearson coefficient of -0.81 and p-value of 0.008. This effect is well known to depend on the jet activity (the higher the jet flux, the lower the equivalent width), and has been observed in other sources \citep[][]{Corbett2000,Foschini12b}. In particular, it is interesting to compare what we observed with what was seen during a flare of the the FSRQ 3C 345 in which the emission lines reacted to a significant flux increase in the continuum \citep{Berton18b}. In that case, the source was deviating from the best-fit relation between the continuum flux and the equivalent width, and that was interpreted as a sign of prominent disk contribution to the typically jet-dominated continuum. In this case, we do not observe any significant deviation during the flare. Therefore, it seems like the jet emission is dominating over the disk at all epochs. \par 
The flux of H$\beta$ may be responding to the continuum variation. The two spectra showing the maximum continuum flux and H$\beta$ flux were observed 29 days apart. As reported in Table~\ref{tab:parameters}, the BLR size is $\sim$15.2 light days. Assuming that this distance is akin to the light crossing time from the central engine to the BLR clouds, it is reasonable to assume that the enhanced H$\beta$ flux we observe in the last spectrum is the echo of the enhanced continuum produced in the nucleus during the flare. However, while the continuum increased by a factor of $\sim$4, the line flux did so only by a factor of $\sim$1.3. Due to the sparse sampling of our observations, neither of these measurements probably reflect the behavior of the source during the flare, but they can still provide some insights. \citet{Cracco16}, in a large sample of NLS1s, found that the continuum and H$\beta$ luminosity are related as L(H$\beta$) $\propto$ $\lambda L(5100)^{1.203}$. If this relation applies to our source as well, the ionizing continuum should have increased only by a factor of 1.24 to account for the observed variation in H$\beta$, much lower than what we actually observe. This may indicate that the vast majority of the flux increase was due to an enhanced activity of the relativistic jet, while the accretion disk, which produces most of the ionizing photons and later affects H$\beta$, played a significantly smaller, although non-negligible, role. This may reflect what is seen in the X-ray spectrum of this source, which is dominated by a power-law component coming from the relativistic jet, while the thermal Comptonization model coming from the disk corona accounts for only 2\% of the total flux \citep{Gallo06a, Kreikenbohm16, Berton19b}. It is worth mentioning that \citet{Gokus21} did not detect any component associated with the accretion disk in their analysis of PKS 2004-447 X-ray spectrum observed after the flare, and this is in good agreement with our conclusions. If the jet flux is strongly enhanced, while the corona contribution does not change significantly, the latter will likely become too weak to be detected. On the opposite, as the soft X-ray excess is associated to a high X-ray flux, hence to high jet activity, it is likely to be the high-energy tail of the synchrotron emission, as it happens for low-frequency peaked BL Lac objects \citep{Foschini09,Foschini20}.
\par

\begin{table}[]
\caption{Spectral measurements at different epochs.}
\label{tab:variability}
    \centering
    \begin{tabular}{l c c c}
    \hline
    Date & F$_{\rm cont}$ & F$_{H\beta}$ & EQW \\
    \hline\hline
    1984-05-02 & 6.70$\pm$0.51 & 3.94$\pm$0.15 & 70.1$\pm$7.0 \\
    2015-10-17 & 5.84$\pm$2.98 & 3.07$\pm$0.54 & 58.5$\pm$5.9 \\
    2015-10-22 & 5.68$\pm$0.87 & 3.14$\pm$0.14 & 69.8$\pm$7.0 \\
    2015-11-09 & 6.25$\pm$0.53 & 3.18$\pm$0.07 & 59.5$\pm$6.0 \\
    2016-03-18 & 9.14$\pm$0.57 & 3.92$\pm$0.13 & 47.8$\pm$4.8 \\
    2016-03-29 & 5.31$\pm$2.45 & 3.41$\pm$0.43 & 103.4$\pm$10.3 \\
    2016-04-02 & 7.16$\pm$0.51 & 3.85$\pm$0.08 & 69.7$\pm$7.0 \\
    2019-04-16 & 6.40$\pm$0.89 & 3.36$\pm$0.13 & 65.4$\pm$6.5 \\
    2019-10-31 & 29.85$\pm$2.92 & 3.59$\pm$0.53 & 20.0$\pm$2.0 \\
    2019-11-29 & 6.56$\pm$1.75 & 4.58$\pm$0.24 & 82.3$\pm$8.2 \\
    \hline
    \end{tabular}
\tablefoot{Columns: (1) Observation date; (2) continuum flux at 5100\AA\ in units of $10^{-17}$ \ergs cm$^{-2}$ \AA$^{-1}$; (3) H$\beta$ integrated flux in units of $10^{-15}$ \ergs cm$^{-2}$; (4) equivalent width of H$\beta$ in \AA. We define it as positive for emission lines. }
\end{table}

\subsection{Eddington ratio}
\begin{figure}[!t]
    \centering
    \includegraphics[width=\hsize]{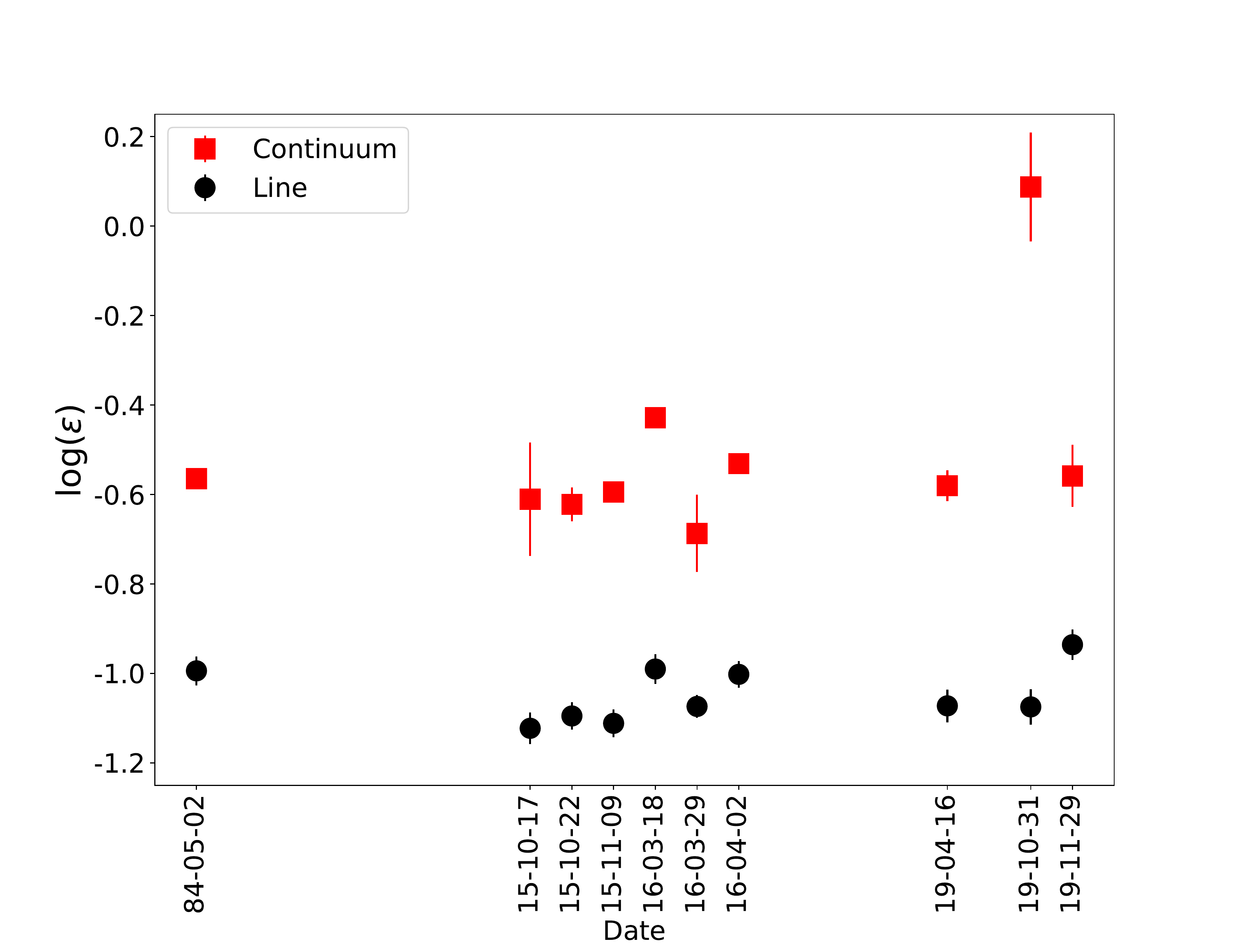}
    \caption{Evolution of the Eddington ratio estimated by using the continuum (red squares) and the H$\beta$ line luminosity (black circles).}
    \label{fig:edd}
\end{figure}
The Eddington ratio is defined as
\begin{equation}
    \epsilon = \frac{L_{\rm bol}}{L_{\rm Edd}} = \frac{L_{\rm bol}}{1.3\times10^{38} \, M_{\rm BH}/M_\odot} \; ,
\end{equation}
that is the ratio between the bolometric luminosity $L_{\rm bol}$ and the Eddington luminosity $L_{\rm Edd}$. This parameter is often believed to be the driver of the quasar MS. Sources with high Eddington ratio show the most prominent Fe II multiplets and the narrowest H$\beta$, while low Eddington sources are characterized by large FWHM(H$\beta$) and little or no Fe II. In NLS1s, the Eddington ratio is typically between 0.1 and 1 \citep{Boroson92, Williams02, Williams04, Grupe10, Xu12}, but even super-Eddington accretion can be observed in a handful of sources \citep{Chen18}. The bolometric luminosity is usually roughly estimated by adopting a simple linear relation that connects it to the continuum luminosity at 5100\AA, $L_{\rm bol}$ = 9$\lambda L_\lambda$(5100 \AA) \citep{Kaspi00}. However, in the light of what was shown in Sect.~\ref{sec:lines}, this approach can be misleading in jetted sources, especially in those like PKS 2004-447 where the jet component dominates the continuum emission. An alternative way is using instead the emission lines luminosity as a proxy for the disk luminosity. As shown in Equation~\ref{eq:greene}, the H$\beta$ luminosity has a linear relation in the log-log plane with the BLR radius. Assuming a photoionization regime, the latter directly depends on the disk luminosity, as
\begin{equation}
    \frac{R_{\rm BLR}}{10^{17} \, \rm cm} = \sqrt{\frac{L_{\rm disk}}{10^{45}\, \rm erg\, s^{-1}}} \; ,
    \label{eq:disk}
\end{equation}
\citep{Koratkar91, Ghisellini09}. The median disk luminosity we obtain from this relation is $2.0\times10^{44}$ \ergs. Under the reasonable hypothesis that the disk luminosity is comparable to the bolometric luminosity without the jet emission, we can use it to estimate the Eddington ratio. \par
The results of both techniques are shown in Fig.~\ref{fig:edd}. On the y-axis a logarithmic scale was adopted to enhance the visibility of the variations. Regardless of the method we used, the Eddington ratio lies within the typical range of NLS1s, with median values 0.29 and 0.09 for continuum- and H$\beta$-based estimate, respectively. While the continuum-based estimate varies from 0.22 on 2015-10-22 to 1.23 during the 2019 flare, the H$\beta$-based estimate has a more limited range, going from 0.07 on 2015-10-17 to 0.11 in the last spectrum. It is worth noting that the estimate based on the H$\beta$ luminosity allows us to measure the Eddington ratio as it was 15.2 days ago, i.e. the BLR light crossing time, before the observation date. Considering this, both estimates agree that the Eddington ratio increased during the flare. This may suggest a possible connection between the accretion disk and the jet, as seen in other AGN flares \citep[e.g.,][]{Grandi04}. 

\subsection{The 2019 flare}
\label{flare19}

\begin{table}[!t]
\caption{Physical parameters of PKS 2004-447 derived from the spectrum of Fig.~\ref{fig:pks_fors}.}
\label{tab:parameters}
    \centering
    \begin{tabular}{l c}
    \hline
    Parameter & Value \\
    \hline\hline
    $M_{\rm BH}$ & $1.5\times10^7$ M$_\odot$ \\
    $r_{\rm g}$ & $2.2\times10^{12}$ cm\\
    $\epsilon$(L$_{5100}$) & 0.27 \\  
    $\epsilon$(H$\beta$) & 0.08 \\
    $L_{\rm disk}$ & $1.6\times10^{44}$ \ergs \\
    $R_{\rm BLR}$ & 15.2 l.d. \\
    $R_{\rm sub}$ & 0.2 pc \\
    $R_{\rm out}$ & 4.7 pc \\
    $R_{\rm NLR}$ & 2450 pc \\    
    \hline
    \end{tabular}
\tablefoot{Columns: (1) Parameter measured. (2) Estimate and measure units of each parameters. }
\end{table}

By using the lines and continuum luminosities of Fig.~\ref{fig:pks_fors}, we tried to derive the scales involved in the inner structure of PKS 2004-447. The results are reported in Table~\ref{tab:parameters}. The BLR radius was derived using Equation~\ref{eq:greene}, and the disk luminosity following Equation~\ref{eq:disk}. For the dust sublimation radius and the outer radius of the torus we used the scaling relations between these quantities and the disk luminosity \citep{Elitzur08}, while to estimate the maximum extension of the NLR we used its correlation with the [O~III] luminosity \citep{Fischer18}. All the values we found are comparable to those derived in the other $\gamma$-ray emitting NLS1 1H 0323+342 \citep{Foschini19}, suggesting that their inner structure is rather similar. \par
These values can provide us an idea of the region where the $\gamma$-ray photons were produced during the 2019 flare. It is known that the minimum Doppler factor needed to account for a source variability is
\begin{equation}
    \delta > \frac{r (1 + z)}{c\tau} \; ,
\end{equation}
where $r$ is the radius of the emitting region, and $\tau$ the observed time scale. Using the 6-hour binned $\gamma$-ray light curve derived by \citet{Gokus21}, the variability between adjacent points indicated that the doubling time is of the order of $\sim 2-4$ hours, consistent with the values observed at hard X-rays by \cite{Berton19b}. If we assume that the jet structure is self-similar, and that its semi-opening angle is 0.1 radians, the size of the emitting region depends on the distance of the dissipation region from the central source. Therefore, if the production of $\gamma$-rays is located relatively close to the jet base at $\sim 10^3 r_{\rm g}$ \citep{Ghisellini10}, the minimum Doppler factor is $\delta \geq 0.6-1.3$, which is a rather weak constraint that can be achieved also with a large viewing angle (\citealp{Schulz15} estimated $\theta < 50^{\circ}$). This is consistent with what is typically observed high-energy emitting radio galaxies (e.g. NGC 6251, $\theta < 47^{\circ}$, $\delta \sim 3.2$, or Cen A, $\theta < 80^{\circ}$, $\delta\sim 1.2$, see \citealp{Chiaberge2001,Chiaberge03, Foschini05}). Furthermore, the $\gamma$-ray spectrum they measured is rather steep. If the production of $\gamma$-rays had occurred close to the black hole, we would indeed expect to see a rather steep spectrum, because of absorption of the most energetic photons due to the BLR gas \citep[e.g.,][]{Romano2020}. 
\citet{Gokus21} argued that the dissipation region is instead close to the molecular torus, but to calculate its radius they assumed a disk luminosity which is one order of magnitude lower than what we derived from H$\beta$. Using our estimate for the torus inner radius, that is the dust sublimation radius and is based on the observed disk luminosity, the minimum Doppler factor required would be
$\delta > 177-354$, which are clearly unrealistic. Therefore, we believe that a dissipation region closer to the central source is, in this case, more likely. \par 


\section{Discussion}

\subsection{Optical classification}
Our new, high-quality spectral data can finally confirm with certainty the classification of PKS 2004-447 as an NLS1, as its spectrum respects all the fundamental criteria. As calculated in \ref{sec:hb}, FWHM(H$\beta) = 1617$ \kms, lower than the 2000 \kms\ threshold. The ratio R5007 = F([O~III])/F(H$\beta$) = 1.53, also complies with the R5007$<$3 limit. Finally, Fe II multiplets are definitely present and rather strong. Specifically, the value of R4570 = $1.14 \pm 0.14$ we found is above the median value of 0.49 found for NLS1s \citep{Cracco16}, but well in agreement with what is often observed in these objects and, in general, in population A sources. We remark here that the 2000 \kms\ threshold is an artificially imposed value that does not reflect any physical difference between sources above and below it. As long as they can be classified as type 1 AGN, all sources within population A roughly share the same physical properties \citep{Marziani18a}. \par 
The H$\beta$ profile of PKS 2004-447, when reproduced with a single function, is better described by a Lorentzian profile. However, the best fit is obtained with three Gaussian functions, one representing the narrow component, and two reproducing the broad component. This result supports the view that the BLR is not homogeneous, but rather stratified depending on its chemical composition \citep{Peterson99, Kovacevic10}. 
This double Gaussian approach to model the BLR can also reproduce the H$\alpha$ profile better than a Lorentzian profile, possibly indicating that the structure of the region where the Balmer lines are produced is similar. As we already pointed out, H$\beta$ may have a slight redward asymmetry that is not seen in H$\alpha$, but only better data will allow us to determine whether this difference is real or just a residual left after the Fe II multiplets subtraction. \par
Another noteworthy fact that can be derived from the optical spectrum is the remarkable width of the forbidden lines. All of them have FWHM of the order of 900 \kms\ after correcting for instrumental resolution, which is significantly broader than what is typically seen in AGN \citep{Peterson}. Assuming that, as hypothesized by \citet{Nelson96}, there is a correlation between the forbidden lines width and the stellar velocity dispersion $\sigma_*$, we can derive $\sigma_* \sim$330 \kms, using its scaling relation with the FWHM([O~II]) derived by \citet{Greene05}. Such a value would be high even for a large elliptical galaxy \citep{Forbes99}, but the host galaxy of PKS 2004-447 is a spiral galaxy with pseudobulge \citep{Kotilainen16}. This suggests that, at least in our source, the forbidden lines width does not obviously correlate with the stellar velocity dispersion. A possible alternative origin for the observed large width may instead be interaction through shocks between the NLR gas and the relativistic jet. Shocks can indeed produce turbulent motion in the gas, which in turn causes the lines to become broader \citep[e.g.,][]{Whittle85, Nesvadba08, Morganti17}. The presence of jet/NLR interaction may also explain the rather high temperature we have derived in Sect.~\ref{sec:temden}, which is above the typical value observed in AGN \citep{Osterbrocklibro}. Further investigation, especially by means of integral-field spectroscopy, can clarify this aspect. 

\subsection{High mass or low mass?}
We have verified that the black hole mass of PKS 2004-447 derived from the optical spectrum is well within the typical range of NLS1s \citep{Peterson11}. However, it is interesting to compare our value with other results in the literature. In particular, we tried to verify how the $f$ factor changes if we use mass indicators which are independent of the inclination. For instance, \citet{Kotilainen16} used the bulge infrared magnitude to estimate a black hole mass of $9\times$10$^7$ M$_\odot$, higher than the value we derived. If we assume that this host-based value is the correct one, and using the weighted average of all our $\sigma$- and FWHM-based measurements, we get $f_\sigma=24.7$ and $f_{\rm FWHM}=8.5$. Both these values are higher than the average correction needed for type 1 AGN, and possibly suggest that inclination and BLR geometry do play a significant role in this source. \par
Another value that can be found in the literature is derived from spectropolarimetric observations of PKS 2004-447 \citep{Baldi16}. These observations are based on the concept that spectropolarimetry can offer a periscopic view of the nucleus, that in this case should be seen edge-on in polarized light. The mass they obtained is much higher than both our estimate and that derived from the host galaxy, $6\times$10$^8$ M$_\odot$. In this case, the scaling factors should be $f_\sigma=165.0$ and $f_{\rm FWHM}=56.7$, which are both extremely high. It is however worth noting that similar observations of other sources revealed that PKS 2004-447 seems to be a unique case among type 1 AGN \citep{Capetti21}. Indeed, even among their sources, PKS 2004-447 is the only one which would require an unrealistically high value for the scaling factor. This could possibly indicate that the polarization results obtained for PKS 2004-447 are not fully reliable, likely because of the insufficient statistics. More observing time is required to strengthen or to reject these hypotheses. \par
In fact, there are multiple powerful physics-based arguments against such high mass values in the class of jetted NLS1s. First, as argued by \citet{Komossa06}, if NLS1s as a class had the highest inclination correction factors and therefore much higher black hole masses, than they should show a much higher fraction of beamed systems than comparison samples of broad-line Seyfert 1s (BLS1s). However, the opposite is the case: beaming and radio loudness is systematically less frequent in NLS1s as a class, than in BLS1s. A second argument was given by \citet{Foschini17}. According to the theory of the blazar sequence \citep{Fossati98, Ghisellini98, Ghisellini16}, both classical blazar classes FSRQs and BL Lacs have a double-humped spectral energy distribution (SED). The first hump originates from synchrotron radiation, while the second one from inverse Compton. Essentially, the blazar sequence can be interpreted in terms of electron cooling and nuclear environments. FSRQs have a photon- and gas-rich environment, and the electron cooling process occurs efficiently via inverse Compton, in particular external Compton, where the seed photons come from the accretion disk, the torus, or the ISM. BL Lacs instead are the opposite. Their environment is photon-starved, and the only cooling mechanism is synchrotron self-Compton. The SED of $\gamma$-NLS1s has a very similar shape to that of FSRQs, but the total emitted power is much lower (cf Fig. 1 in \citealp{Foschini10}). This is the case also for PKS 2004-447 \citep[see][their Fig.~5, left panel]{Paliya13}. This is in agreement with the findings of \citet{Foschini15}, who calculated the jet power of NLS1s, FSRQs, and BL Lacs. They found that jetted NLS1s have systematically lower jet power than FSRQs, but comparable to BL Lacs. This can be interpreted in terms of mass difference since, as predicted theoretically by \citet{Heinz03}, the jet power scales non linearly with the black hole mass. The physics behind jetted NLS1s and FSRQs is exactly the same, hence the similar SED, but NLS1s are scaled-down versions of FSRQs because of their lower black hole mass. This result is confirmed by the observations of their radio luminosity function, that clearly showed how jetted NLS1s are the low-luminosity tail of FSRQs \citep{Berton16c}. Vice versa, in BL Lacs the mass is similar to FSRQs, but the cooling mechanism is different because of the different environment, and this translates into the lower jet power and the different SED. It is worth noting that, if NLS1s had high-mass black holes, they would have a low jet power in a high-density environment. This would indicate that the relativistic electrons of their jets are not cooling despite the photon-rich environment, and this is clearly unphysical \citep{Foschini17}. Therefore, the only reasonable explanation for the different jet power in FSRQs and NLS1s is the black hole mass, and NLS1s do not fit in the classical blazar sequence because their behavior is not regulated only by electron cooling. The blazar sequence is missing an ingredient, that is the black hole mass or, if NLS1s are the progenitors of FSRQs, the evolution of AGN \citep{Berton17}. 

\subsection{A rare $\gamma$-ray NLS1/CSS hybrid}
We confirmed that the optical spectrum of PKS 2004-447 is definitely that of a rather typical NLS1s. The spectral energy distribution of PKS 2004-447, along with its $\gamma$-ray emission, confirm its blazar-like nature \citep{Foschini09, Abdo09c, Paliya13}. Also the one-sided morphology of the relativistic jets and the high brightness temperature show that relativistic beaming in this source is not negligible \citep{Schulz15}. Its radio properties, however, clearly indicate the source is also a CSS of relatively low luminosity, with a turn over below 1~GHz \citep{Schulz15}. This result is not new as, historically, PKS 2004-447 was the first source identified as a potential link between NLS1s and CSS \citep{Oshlack01, Gallo06a}. The detection of $\gamma$-ray emission from a CSS is instead quite rare, as only five objects with this classification have been included in the 4FGL catalog, along with six more classified as young radio galaxies \citep{Abdollahi20}. \par
In the case of PKS 2004-447, possible emission from a counter-jet has been found in VLBA observations at 1.5~GHz \citep{Schulz15}. If this is the case, the jet of PKS 2004-447 may have a non-negligible inclination with respect to the line of sight ($\theta < 50^\circ$). The minimum Doppler factor we have estimated from the observed variability at X- and $\gamma$ rays \citep{Berton19c,Gokus21} (see Sect.~\ref{flare19}), requires that the dissipation occurs close to the black hole. This is also consistent with what is observed in some $\gamma$-ray emitting radio galaxies \citep{Foschini05}. \par
An alternative option is that the source of $\gamma$-ray emission is not located close to the black hole, but farther away from it. In their VLBA map at 1.5~GHz, \citet{Schulz15} found signs of coherent emission coming from a hotspot $\sim$45 mas away from the nucleus (projected size $\sim$170 pc), which corresponds to a change in the jet position angle. This potentially indicates ongoing interaction between the relativistic jet plasma and the ISM. However, according to the jet model by \cite{Blandford79}, to have a dominant contribution from curvature, ultrarelativistic speed and small viewing angle are required. This is not verified in the present case, as clearly shown by radio maps. \par
Another interesting possibility is that the inclination of the relativistic jet in the core is different from that at larger scales. If the jet axis has changed its position with time, it is possible that in the past it had a large inclination with the line of sight, while nowadays it is closely aligned with it. An episode of jet realignment from radio galaxy to blazar has already been observed directly \citep{Hernandezgarcia17}, and precession of the jet axis was also invoked to explain the properties of the jetted NLS1 Mrk 783 \citep{Congiu17, Congiu20}. \par
PKS 2004-447 is not the only example of CSS/NLS1 with $\gamma$-ray emission, as the strong radio source 3C 286 can also be classified as an NLS1 \citep{Berton17, Liao20, Yao21}, and it is included in the Fermi catalog \citep{Abdollahi20}. For 3C 286, radio observations were able to obtain a rather high inclination value of $\sim48^\circ$, and alternative mechanisms to produce its $\gamma$-ray emission such as jet/ISM interaction have been suggested \citep{An17}. However, a number of jetted NLS1s do actually show a compact morphology at kpc-scale and a steep radio spectrum \citep{Berton18a}. Most of them, however, show lower luminosities when compared to typical CSS, and they mostly belong to the class of low-luminosity compact sources \citep{Kunert10a}. Given all these similarities between NLS1s and CSS, it has been suggested that some CSS, especially those of relatively low luminosity and with high-excitation emission lines, may represent the parent population of $\gamma$-ray NLS1s \citep{Berton16c}. In other words, some of them may be the exact same kind of source seen at different angles. This tentative unification has not been fully confirmed yet, but the existence of hybrid sources as PKS 2004-447 may be an indication that this idea is at least in part correct \citep[see also Sec.~7.1.3 of][]{Odea21}. \par
It is also worth noting that at least a fraction of luminous CSS are also characterised by moderate to high Eddington ratios typical of Population A sources \citep{Wu09}. For instance, the high R4570 and other properties of 3C 57 suggest that this luminous CSS source is optically young (or more likely rejuvenated) in addition to being radio young, as CSS have been suggested to be \citep[e.g.,][]{Fanti11}. Considering that luminous CSS have relatively large mass black holes, it seems at least conceivable that they include already massive black holes and be rather evolved sources, rejuvenated by a new accretion episode yielding a moderate-to-high Eddington ratio. Conversely, low-luminosity CSS may be their younger counterparts at their first accretion episodes, and could be directly linked to NLS1s. 


\section{Summary}
In this paper we analyzed a set of optical spectra of the $\gamma$-ray emitting source PKS 2004-447. Thanks to our new data, we can confirm its classification as NLS1 galaxy, belonging to population A3 of the quasar main sequence. From a detailed spectral analysis, we found that the source probably has some absorption along the line of sight, although this result is not consistent with past X-ray observations. The temperature and density values we derived, along with the width of the narrow lines, may suggest that some interaction between the NLR gas and the relativistic jet is present. The black hole mass we derived from the optical spectrum using the H$\beta$ line is $(1.5\pm0.2)\times10^7$ M$_\odot$. Despite the uncertainties related to this estimate, several arguments allow us to rule out that PKS 2004-447, and NLS1s in general, are powered by black holes with very large mass. \par
The long-term variability of the spectra indicates that the source continuum emission is rather stable, with the noteworthy exception of the flare, seen also in $\gamma$- and X-rays, of October 2019. We do not know how and if the H$\beta$ line has been responding to the flare, since the reverberation time is shorter than the gap between our observations, but we can speculate that most of the continuum variation is produced by the relativistic jet and not by the accretion disk. 
The Eddington ratio is within the typical range of NLS1s (0.09-0.29, depending on how it is measured), and we found a significant increase during the 2019 flare.   \par
We finally discussed the role of PKS 2004-447 in the unification between NLS1s and CSS. The radio properties of our source clearly suggest that it belongs to both classes, and that its relativistic jet may have a relatively high inclination with respect to the line of sight, given the possible detection of its counterjet in past radio observations. PKS 2004-447 is then the second hybrid CSS/NLS1 with $\gamma$-ray emission, after 3C 286. Continuous multiwavelength monitoring of this object, particularly by means of optical spectroscopy, may help us to finally solve the puzzle of NLS1s and find their link with CSS and other young radio galaxies.

\begin{acknowledgements}
The authors are grateful to the anonymous referee for the constructive comments. 
M.B, S.C., and L.F. acknowledge the financial support from the visitor and mobility program of the Finnish Centre for Astronomy with ESO (FINCA), funded by the Academy of Finland grant nr. 306531.
E.C. acknowledges support from ANID project Basal AFB-170002. 
J.K. acknowledges financial support from the Academy of Finland, grant 311438.
I.B. acknowledges Kulttuurirahasto for their continuing support.
Based on observations made with ESO Telescopes at the La Silla Paranal Observatory under programmes ID 096.B-0256 and 0104.B-0587. 
This paper includes data gathered with the 2.5 meter Dupont Telescope located at Las Campanas Observatory, Chile. 
This research has made use of the NASA/IPAC Extragalactic Database (NED) which is operated by the Jet Propulsion Laboratory, California Institute of Technology, under contract with the National Aeronautics and  Space Administration. 
\end{acknowledgements}

\bibliographystyle{aa}
\bibliography{./biblio}

%

\end{document}